\documentclass[10pt, prb, aps, twocolumn, showpacs, citeautoscript, floatfix, reprint, amsmath, amssymb, notitlepage, superscriptaddress]{revtex4-1}

\usepackage{graphicx}
\usepackage{stmaryrd}
\usepackage{rotating}
\usepackage{amsmath}
\usepackage{amsfonts}
\usepackage{amssymb}
\usepackage[countmax]{subfloat}
\usepackage{dcolumn} % align table columns on decimal point
\usepackage{bm} % bold math
\usepackage{color}

\usepackage{float}
\usepackage{latexsym}

\begin{document}

\title{Absence of equilibrium edge currents in theoretical models of topological insulators}

%\title{Reviving the absent edge currents in topological insulators}

%\title{Mechanisms to revive the absent edge currents in topological insulators}

%\title{Absence and revival mechanisms for edge currents in topological insulators}

%\title{Absence of edge currents in topological insulators and mechanisms to revive them}

%\title{Absence and revival of edge currents in topological insulators}

%\title{Absence of edge currents in topological insulators and their revival by gate voltage}

%\title{Absence of edge currents in topological insulators and how to revive them by gate voltage}

%\title{Electrically controllable quantized edge currents in topological quantum dots}

%\title{Electrically controllable nonlocal edge currents in topological insulators}

\author{Wei Chen}

\affiliation{Department of Physics, PUC-Rio, 22451-900 Rio de Janeiro, Brazil}

\date{\today}

\begin{abstract}

The low energy sector of 2D and 3D topological insulators (TIs) exhibits propagating edge states, which has speculated the existence of equilibrium edge currents or edge spin currents. We demonstrate that if the low energy sector of TIs is regularized in a straightforward manner into a square or cubic lattice, then the current from the edge states is in fact canceled out exactly by that from the valence bands, rendering no edge current. This result serves as a warning that for any equilibrium property of topological insulators, the contribution from the valence bands should not be overlooked. In these regularized lattice model, there is a finite edge current only if the Dirac point of the edge states is shifted away from the chemical potential, for instance by doping, impurities, edge confining potential, surface band bending, or gate voltage. The edge current in small quantum dots as a function of the gate voltage is quantized, and the edge current can flow out of the gated region up to the decay length of the edge state. 

\end{abstract}

\maketitle

\section{Introduction}

The existence of edge states represents one of the defining properties of topological insulators (TIs)\cite{Hasan10,Qi11,Bernevig13}. Particularly for TIs realized in two- (2D) and three-dimension (3D), the edge states manifest as $D-1$ dimensional Dirac cones at the low energy sector of the band structure. The wave functions of these states can be calculated by projecting the low energy Dirac Hamiltonian into a semi-infinite half space\cite{Jackiw76,Hasan10,Qi11,Bernevig13}, which yields wave functions localizing at the edges with a decay length given by the Fermi velocity divided by the bulk gap $\xi=\hbar v_{F}/M$, and along the edge they are propagating states with a finite group velocity. Depending on the symmetry of the TI\cite{Schnyder08,Ryu10,Chiu16}, the edge states may be either spinless states circulating the boundary with a finite group velocity, such as in 2D class A, or spinful states with spin up circulating the boundary in one direction and spin down in the opposite direction, such as the situation of class AII systems in 2D and 3D. Because of the way they circulate the boundary, the edge states naturally speculate the existence of an equilibrium edge current in TIs\cite{Buttiker09,Sonin11,Ando13,Maekawa17}, which would be an edge charge current for class A, and an edge spin current for class AII. Pushing this speculation further, the circulating edge states suggest that the edge currents at the opposite edges must flow in opposite directions, and the directions of the currents should be independent from the chemical potential $\mu$, since the direction of circulation does not depend on $\mu$.

In this article, we demonstrate that the above naive expectation for the existence of edge currents has a serious flaw, namely it does not include the contribution from the valence bands. If the TI is geometrically confined between two boundaries, i.e., has a finite width in 2D or a finite thickness in 3D, then the band structure consists of multiple bands, with the number of bands determined by the width or thickness. This often corresponds to experimental situations, such as 2D TI made of quantum wells of finite width\cite{Konig07,Konig08}, and 3D TI thin films of few quintuple layers thickness\cite{Hasan11,Ando13,He13,Tian17}. We elaborate numerically that if the low energy Dirac cone is regularized into a lattice model in a straightforward manner and the lattice is geometrically confined, then the valence bands of TI also contribute to an edge current which exactly cancels out that from the edge states, rendering no net edge current. The statement remains valid regardless the temperatures and the distance between the edges, as well as the choice of the band parameters within the regularized lattice model. Our investigation covers a variety of well-known 2D and 3D models of TIs belonging to different symmetry classes, which suggests that these statements are relevant to a great majority of TIs in reality. Especially, our result prompts the caution that the valence bands should not be overlooked when discussing the equilibrium properties of TIs.

The cancellation from the valence bands also indicates that there is a finite edge current if the Dirac point is shifted away from the chemical potential $\mu$, since the cancellation would not be complete in this case. It then follows that the direction of the current in fact depends on whether the chemical potential lies above or below the Dirac point. This suggests a finite edge current can be induced by various mechanisms that shifts the chemical potential in reality, such as doping\cite{Hsieh09,Zhang11,Kondou16}, surface band bending\cite{Bahramy12}, or surface inhomogeneity\cite{Beidenkopf11}, and edge confining potential\cite{Akhmerov08}. In particular, we will investigate gating the TI, which leads to an electrically controllable edge current. For TIs geometrically confined in a small lattice\cite{Cho12,Claro19,Jing19,Huang19}, we further uncover that the edge current is geometrically quantized as a function of the gate voltage. Moreover, if the small lattice is only partially gated, then the edge current can extend to the ungated region up to the decay length of the edge state, which can serve as a nonlocal dissipationless current or spin current generator. Finally, although these equilibrium edge currents do not show up in nonequilibrium transport measurements, experiments based on measuring the magnetic or electric field they produce will be proposed.

\section{Lattice models for topological insulators}

\subsection{Regularization of TIs on square or cubic lattices}

Our purpose is to investigate the equilibrium currents in TIs produced by the entire band structure, including both the edge state Dirac cone and the bulk bands, by means of minimal lattice models that can take into account these features. For this purpose we start from the low energy effective Hamiltonian of 2D and 3D TIs described by a linear Dirac model 
\begin{eqnarray}
H=\sum_{\ell=1}^{D}A\,k_{\ell}\Gamma^{\ell}+M\,\Gamma^{0},
\end{eqnarray}
where the $\Gamma^{\ell}$ matrices satisfy the Clifford algebra $\left\{\Gamma^{\ell},\Gamma^{m}\right\}=2\delta_{\ell m}$, whose precise forms depend on the symmetry classes\cite{Schnyder08,Ryu10,Chiu16,Chen19_universality_class}. The kinetic terms $k_{\ell}$ signifies the usual linear band crossing, and $M$ is the mass term that controls the topology. We proceed to regularize the Hamiltonian to cover the entire Brillouin zone (BZ) by
\begin{eqnarray}
&&A\,k_{\ell}\rightarrow A\sin k_{\ell}
\nonumber \\
&&M\rightarrow M+B\sum_{\ell=1}^{D}k_{\ell}^{2}
\rightarrow M+4B-2B\sum_{\ell=1}^{D}\cos k_{\ell},
\end{eqnarray}
where the $k^{2}$ term is important to regularize the model on a lattice\cite{Araujo19}.
The basis of the Hamiltonian depends on the symmetry classes, which may be electrons in different orbitals with or without spins\cite{Schnyder08,Ryu10,Chiu16}. We then Fourier transform the second quantized Hamiltonian into real space to construct the 2D square lattice or 3D cubic lattice Hamiltonian. For instance, if the basis are electron creation and annihilation operators $c_{iI\sigma}^{\dag}$ and $c_{iI\sigma}$ with momentum $k$, orbital $I$, and spin $\sigma$, then the Fourier transform leads to
\begin{eqnarray}
&&\sum_{k}\cos k_{\ell}c_{kI\sigma}^{\dag}c_{kJ\sigma}\rightarrow\frac{1}{2}
\sum_{i\sigma}\left\{c_{iI\sigma}^{\dag}c_{i+\ell J\sigma}+c_{i+\ell I\sigma}^{\dag}c_{iJ\sigma}\right\},
\nonumber \\
&&\sum_{k}i\sin k_{\ell}c_{kI\sigma}^{\dag}c_{kJ\sigma}\rightarrow\frac{1}{2}
\sum_{i\sigma}\left\{c_{iI\sigma}^{\dag}c_{i+\ell J\sigma}-c_{i+\ell I\sigma}^{\dag}c_{iJ\sigma}\right\},
\nonumber \\
\end{eqnarray}
and so follows the lattice Hamiltonian defined on sites $i$.

\subsection{2D class A \label{sec:2D_class_A}} 

\begin{figure}[ht]
\begin{center}
\includegraphics[clip=true,width=0.99\columnwidth]{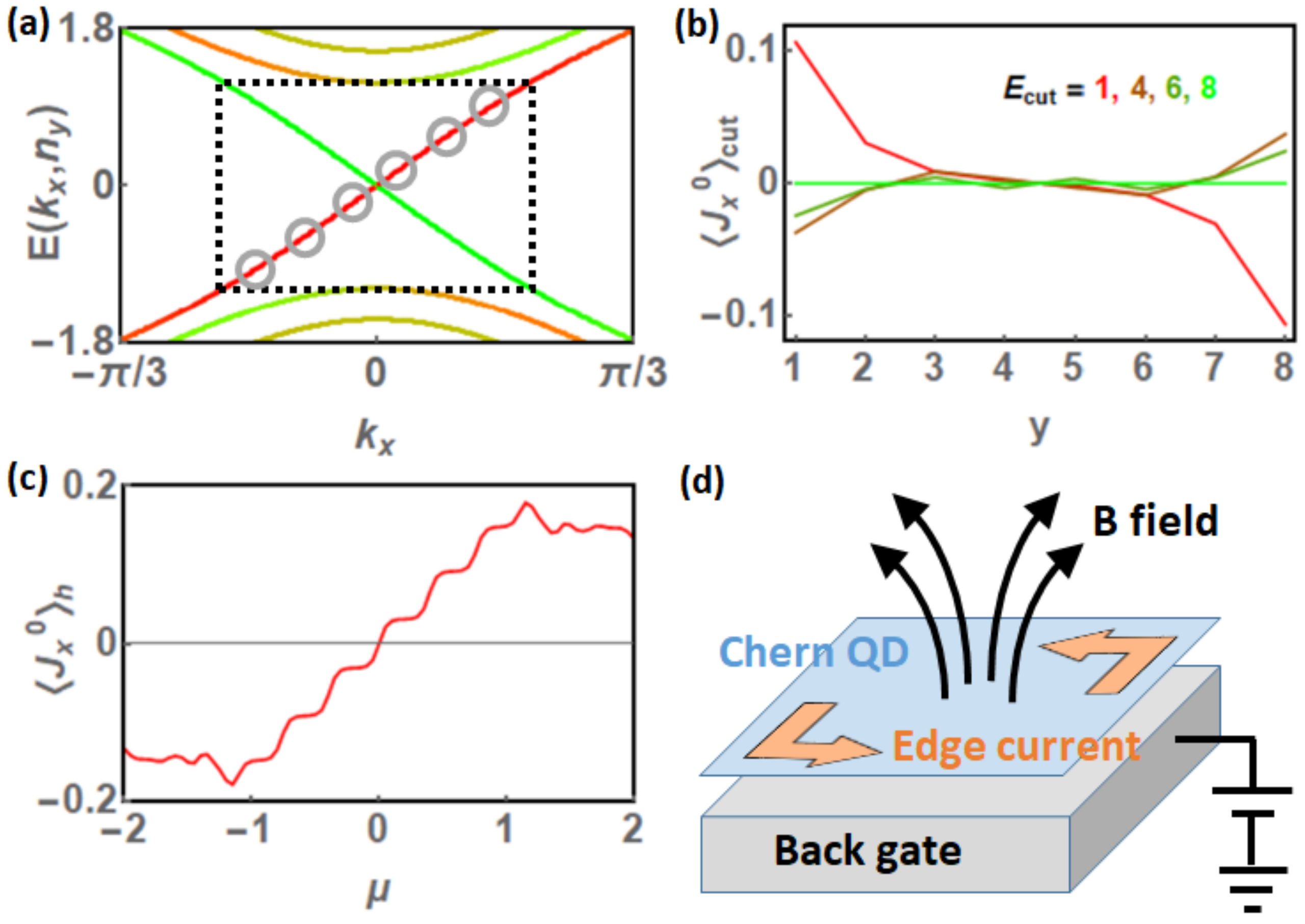}
\caption{(a) The band structure of a strip (PBC in ${\hat{\bf x}}$ and OBC in ${\hat{\bf y}}$) of 2D Chern insulator, with red and green colors indicating the wave function is more localized the $y=1$ or $y=N_{y}$ edge. The grey circles inside the dashed box indicate geometrically quantized edge states in the strip. (b) The expectation value of charge current operator, where the $E_{cut}=1$ case includes only the edge states contribution, and $E_{cut}=8$ includes all the bands. (c) Edge current as a function of the chemical potential in the Chern insulator strip, which shows a stairwise feature in the range $|\mu|<|M|=1$ because of geometric quantization. (d) The proposal of measuring the quantization, where the magnetic flux through a quantum dot of Chern insulator is quantized as a function of the gate voltage. } 
\label{fig:Chern_results}
\end{center}
\end{figure}

The 2D class A models are the simplest ones to demonstrate the absence of edge current in the regularized lattice models introduced in Sec.~II A. We use the prototype Chern insulator to demonstrate our statements in this class. The model is described by the spinless basis $\psi=\left(c_{{\bf k}s},\;c_{{\bf k}p}\right)^{T}$ and Hamiltonian\cite{Qi11,Bernevig13}
\begin{eqnarray}
H({\bf k})&=&A\sin k_{x}\sigma^{x}+A\sin k_{y}\sigma^{y}
\nonumber \\
&+&\left(M+4B-2B\cos k_{x}-2B\cos k_{y}\right)\sigma^{z},
\end{eqnarray} 
which leads to the following lattice model
\begin{eqnarray}
&&H=\sum_{i}t\left\{-ic_{is}^{\dag}c_{i+ap}
+ic_{i+as}^{\dag}c_{ip}+h.c.\right\}
\nonumber \\
&&+\sum_{i}t\left\{-c_{is}^{\dag}c_{i+bp}+c_{i+bs}^{\dag}c_{ip}+h.c.\right\}
\nonumber \\
&&+\sum_{i\delta}t'\left\{-c_{is}^{\dag}c_{i+\delta s}+c_{ip}^{\dag}c_{i+\delta p}+h.c.\right\}
\nonumber \\
&&+\sum_{i}\left(M+4t'\right)\left\{c_{is}^{\dag}c_{is}
-c_{ip}^{\dag}c_{ip}\right\}-\sum_{iI}\mu\,c_{iI}^{\dag}c_{iI},
\label{Hamiltonian_2DclassA}
\end{eqnarray} 
where $t=A/2$, $t'=B$, $I=\left\{s,p\right\}$ are the orbitals, and $\delta=\left\{a,b\right\}$ are the lattice constants along planar directions. Although the chemical potential $\mu$ is frequently ignored because it does not alter the topology of the system, we will elaborate its importance for the existence of the edge current. We impose periodic boundary condition (PBC) in the ${\hat{\bf x}}$ direction and open boundary condition (OBC) in the ${\hat{\bf y}}$ direction, such that the edge currents are localized at the two edges $y=1$ and $y=N_{y}$. The band structure at parameters $t=t'=-M=1$ with a finite width $N_{y}=8$ is shown in Fig.~\ref{fig:Chern_results} (a). For each eigenstate $|k_{x},n_{y}\rangle$ we calculate the weight of its wave function close to the $y=1$ boundary minus that close to the $y=N_{y}$ boundary
\begin{eqnarray}
\tilde{n}_{k_{x},n_{y}}=\left(\sum_{1\leq y\leq N_{y}/2}-\sum_{N_{y}/2+1\leq y\leq N_{y}}\right)|\psi_{k_{x},n_{y}}(y)|^{2}.
\end{eqnarray}
The color in the band structure in Fig.~\ref{fig:Chern_results} (a) indicates a positive (red) or negative (green) $\tilde{n}_{k_{x},n_{y}}$, i.e., whether the $|k_{x},n_{y}\rangle$ is more localized in the $y=1$ or the $y=N_{y}$ boundary. The result clearly indicates that the branch of the Dirac cone with positive group velocity is localized at the $y=1$ edge (red), whereas the branch of negative group velocity is localized at the $y=N_{y}$ edge (green). In addition, the valence bands at $k_{x}>0$ that have negative group velocities are also more localized at the $y=1$ edge, indicating that they contribute to a current {\it against} that of the edge states, and likewise the valence bands at $k_{x}<0$ that have positive group velocity are more localized at $y=N_{y}$ edge and also contribute to a current against that of the edge states.

We proceed to quantify the contribution from the edge states and that from the valence bands by calculating the expectation value of the local current operator, constructed from the equation of motion of the density operator $n_{i}=\sum_{I}c_{iI}^{\dag}c_{iI}$ written in the form of continuity equations 
\begin{eqnarray}
\dot{n}_{i}=\frac{i}{\hbar}\left[H,n_{i}\right]=-{\boldsymbol\nabla}\cdot{\bf J}_{i}^{0}=-\frac{1}{a}\sum_{\delta}\left(J_{i,i+\delta}^{0}+J_{i,i-\delta}^{0}\right),\;\;\;
\label{ndot_commutator}
\end{eqnarray}
which defines the charge current operator $J_{i,i+\delta}^{0}$ running from site $i$ to $i+\delta$, and that run from $i$ to $i-\delta$. We define the currents flowing in the positive bonds as $J_{x}^{0}\equiv J_{i,i+a}^{0}$ and $J_{y}^{0}\equiv J_{i,i+b}^{0}$, which have the expressions
\begin{eqnarray}
J_{x}^{0}&=&\frac{a}{\hbar}t\left\{c_{is}^{\dag}c_{i+ap}
+c_{i+ap}^{\dag}c_{is}+c_{ip}^{\dag}c_{i+as}
+c_{i+as}^{\dag}c_{ip}\right\}
\nonumber \\
&+&\frac{a}{\hbar}it'\left\{-c_{is}^{\dag}c_{i+as}+c_{i+as}^{\dag}c_{is}
+c_{ip}^{\dag}c_{i+ap}-c_{i+ap}^{\dag}c_{ip}\right\}
\nonumber \\
J_{y}^{0}&=&\frac{b}{\hbar}it\left\{-c_{is}^{\dag}c_{i+bp}
+c_{i+bp}^{\dag}c_{is}+c_{ip}^{\dag}c_{i+bs}
-c_{i+bs}^{\dag}c_{ip}\right\}
\nonumber \\
&+&\frac{b}{\hbar}it'\left\{-c_{is}^{\dag}c_{i+bs}+c_{i+bs}^{\dag}c_{is}
+c_{ip}^{\dag}c_{i+bp}-c_{i+bp}^{\dag}c_{ip}\right\}.
\nonumber \\
\end{eqnarray}
To separate the contribution from the Dirac cone and that from the bulk bands, in calculating the expectation value we introduce an energy cutoff $E_{cut}$ within which we sum the eigenstates
\begin{eqnarray}
\langle J_{x}^{0}\rangle_{cut}=\sum_{n}\langle n|J_{x}^{0}|n\rangle f(E_{n})\theta(E_{cut}-|E_{n}|),
\label{current_expectation_Ecut}
\end{eqnarray}
where $E_{n}$ and $|n\rangle$ are eigenenergy and eigenstate of the lattice model, and $f(E_{n})=\left(e^{E_{n}/k_{B}T}+1\right)^{-1}$ is the Fermi distribution, $\theta(E_{cut}-|E_{n}|)$ is the step function that selects the energy window within which the states are included, and we choose $k_{B}T=0.03$  that corresponds to the room temperature throughout the article, assuming the energy unit $t=$ eV.

The results shown in Fig.~\ref{fig:Chern_results} (a) indicates that choosing an energy cutoff smaller than the bulk gap $E_{cut}<|M|=1$, which corresponds to counting the contribution from the Dirac cone alone, indeed yields a finite current localized at the two edges. As increasing $E_{cut}$, the spatial profile of the current starts to alter, and eventually the current vanishes everywhere at a large $E_{cut}=8$ that includes contributions from all the valence bands, proving that the two contributions cancel out exactly. This conclusion holds regardless the temperature, distance between the two edges $N_{y}$, and band parameters $\left\{t,t',M\right\}$ within our regulated lattice model, provided the system remains in the topologically nontrivial phase $M<0$. Thus unless one has certain experimental probe that can discern the edge current contributed only within an energy window near the chemical potential, one cannot resolve the contribution from the edge states alone and the total current should be zero. Remarkably, the edge current profile at a specific choice of energy cutoff $E_{cut}$ (each line in Fig.~\ref{fig:Chern_results} (b)) remains the same at any temperature, even up to a unrealistically high temperature $k_{B}T=1\sim 10000$K. The difference at different temperatures remains less than $10^{-10}$ in our numerical calculation.

It is also immediately clear that if the Dirac point is shifted away from the zero energy by adjusting chemical potential $\mu$, then a finite edge current occurs, since some states below the Dirac cone are omitted if the chemical potential lies below the Dirac point, and some more states are included if the chemical potential lies above the Dirac cone, hence the cancellation from the valence bands is not exact. This argument also implies that the edge current must change sign as chemical potential sweeps through the Dirac point $\mu=0$, as confirmed by the total current close to the $y=1$ edge $\langle J_{x}^{0}\rangle_{h}=\sum_{1\leq y\leq N_{y}/2}\langle J_{x}^{0}\rangle$ shown in Fig.~\ref{fig:Chern_results} (c).

Moreover, Fig.~\ref{fig:Chern_results} (c) also indicates that the edge current as a function of the chemical potential $\mu$ is stepwise quantized within the bulk gap $|\mu|<|M|=1$ (the steps are smeared by temperature). This behavior is due to the fact that the Dirac cone is geometrically quantized by the small size of the strip, as indicated by the grey circles inside the dashed box in Fig.~\ref{fig:Chern_results} (a). As a result, every time $\mu$ sweeps through one of these quantized edge states, the edge current increases by one unit. The number of plateaux $N_{p}$ and the edge current quantum $\Delta J_{x}^{0}$ can be understood in the following manner. The width of the dashed box in Fig.~\ref{fig:Chern_results} (a) is the insulating gap divided by the group velocity of the Dirac cone $k_{p}=M/\hbar v_{F}$, and the spacing between states due to geometric quantization is $\Delta k=2\pi/L_{x}=2\pi/N_{x}a$, so the number of circles that can fit into the dashed box in Fig.~\ref{fig:Chern_results} (a) is 
\begin{eqnarray}
N_{p}=\frac{k_{p}}{\Delta k}=\frac{MN_{x}a}{2\pi\hbar v_{F}}.
\label{number_of_plateaux}
\end{eqnarray} 
On the other hand, because each quantized edge state spreads out the whole edge of length $L_{x}=N_{x}a$ and decays into the bulk with length $\xi$, one may approximate the 2D density of the edge state by $n_{2D}=1/N_{x}a\xi$. Consequently, each quantized edge state contributes to a current density $\Delta j_{2D}^{0}=en_{2D}v_{F}$. The edge current quantum is this value multiplied by the cross section $\xi$ of the current flow 
\begin{eqnarray}
\Delta J_{x}^{0}=\Delta j_{2D}^{0}\xi=\frac{ev_{F}}{N_{x}a}.
\label{height_of_plateaux}
\end{eqnarray}
Equations (\ref{number_of_plateaux}) and (\ref{height_of_plateaux}) well explain the number and height of the plateaux in Fig.~\ref{fig:Chern_results} (c). Finally, we remark that the finite edge current only occurs in the topologically nontrivial phase $M<0$ in this model. In the topologically trivial phase $M>0$ there is no edge state, and neither do the valence bands contribute to a current. Finally, an experimental setup for measuring this quantized edge current is proposed in Fig.~\ref{fig:Chern_results} (d), which will be discussed in the next section for the more realistic quantum dot geometry.

%{\cblue (2) Invoke my flux quantum paper too. But say that this is geometrically quantized, not field-quantized, so there is no electric flux quantum and things like that. }

\begin{figure}[ht]
\begin{center}
\includegraphics[clip=true,width=0.99\columnwidth]{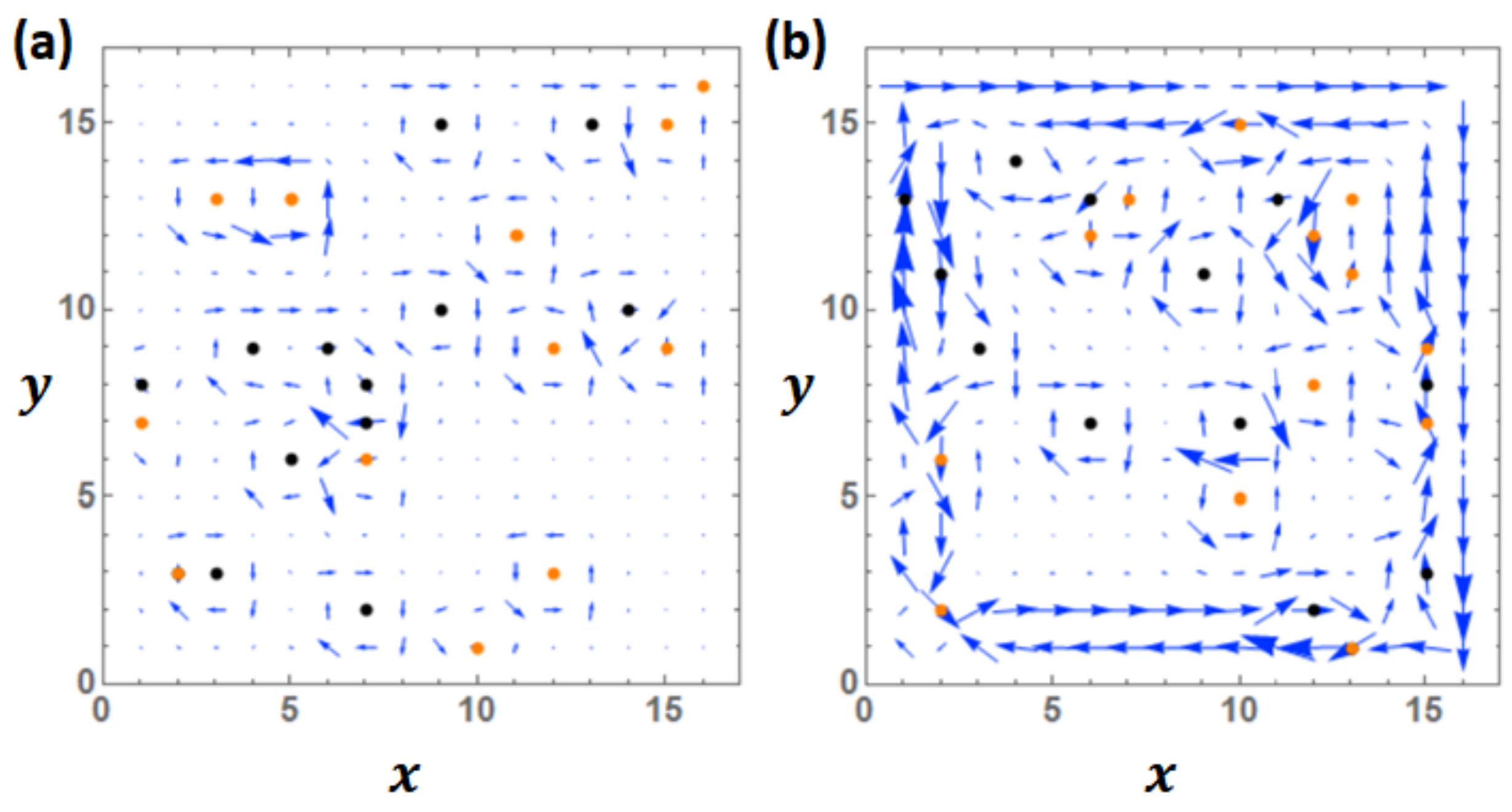}
\caption{(a) Local current in Chern insulators induced by impurities in a $16\times 16$ lattice with impurity potential $U_{imp}=0.3$ (black) and $-0.3$ (orange). The largest arrow has magnitude $|\langle J_{i,i+\delta}^{0}\rangle|=0.016$. (b) Local current at $U_{imp}=0.6$ (black) and $-0.6$ (orange), and in addition a confining potential $V_{con}=0.6$ at the edge sites, where the largest arrow corresponds to $|\langle J_{i,i+\delta}^{0}\rangle|=0.026$.  } 
\label{fig:Chern_2D_impurity_results}
\end{center}
\end{figure}

\subsection{Local current promoted by weak impurities and edge confining potential in a Chern quantum dot \label{sec:impurity_confining}}

The results in Sec.~\ref{sec:2D_class_A} prompts us to explore the effects of weak impurities and edge confining potentials, since it helps to understand the situation when the chemical potential is altered locally. For this purpose, we consider random impurities of density $n_{imp}=N_{imp}/N_{x}N_{y}$ by adding the term
\begin{eqnarray}
H_{imp}=\sum_{i\in imp,I}U_{imp}c_{iI}^{\dag}c_{iI},
\end{eqnarray}
into our Hamiltonian in Eq.~(\ref{Hamiltonian_2DclassA}), where $i\in imp$ denotes impurity sites. In Fig.~\ref{fig:Chern_2D_impurity_results} we present the simulation in an open island of Chern insulator quantum dot (OBC imposed in both ${\hat{\bf x}}$ and ${\hat{\bf y}}$ directions), which confirms that a local current is induced around the impurities. The local current forms a vortex circulating the impurity, and the direction of the vorticity depends on the sign of $U_{imp}$. The vortices interfere if the impurities are getting too close to each other. Comparing small $U_{imp}=\pm 0.3$ and large $U_{imp}=\pm 0.6$, one sees that the magnitude of the local current is enhanced at large impurity potentials, at least within this weak impurity strength we investigate. For the $U_{imp}=\pm 0.6$ case in Fig.~\ref{fig:Chern_2D_impurity_results} (b) we also include a confining potential at the edge sites $i\in edge$
\begin{eqnarray}
H_{con}=\sum_{i\in edge,I}V_{con}c_{iI}^{\dag}c_{iI},
\end{eqnarray}
with $V_{con}=0.6$ chosen the same as the impurity potential. One sees that the confining potential indeed induces an edge current circulating around the edge, since it locally shifts the chemical potential, and the magnitude of the current (size of the arrows in Fig.~\ref{fig:Chern_2D_impurity_results} (b)) is generally larger than that induced by point-like impurities. Interestingly, Fig.~\ref{fig:Chern_2D_impurity_results} (b) also reveals that the current on the boundary sites (e.g., $y=1$) and the sites next to the boundary (e.g., $y=2$) actually have opposite directions of flow, meaning that the edge current induced by the edge confining potential is a laminar flow whose direction of flow depends on the distance to the edge. Note that this confining potential induced edge current bears a striking similarity with the edge current in the quantum Hall effect of 2D electron gas, which is also generated by edge confining potential\cite{Halperin82,Smrcka84,MacDonald84,Streda87,Buttiker88,Chklovskii92}, although there are no Landau levels in our problem.

\begin{figure}[ht]
\begin{center}
\includegraphics[clip=true,width=0.8\columnwidth]{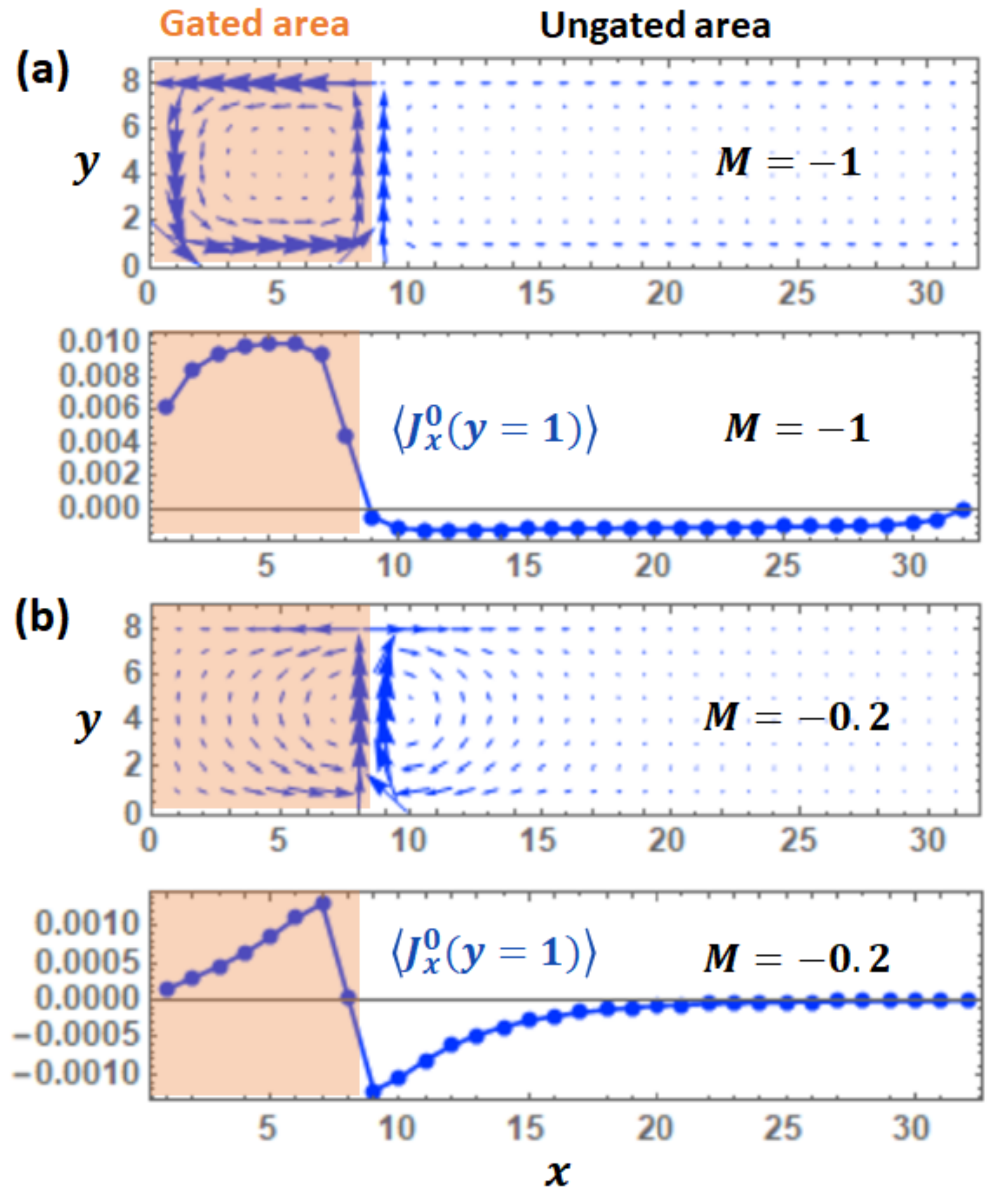}
\caption{(a) (top) The pattern of local current $\langle{\bf J}^{0}\rangle=(\langle J_{x}^{0}\rangle,\langle J_{y}^{0}\rangle)$ in a $32\times 8$ quantum dot at bulk gap $M=-1$, where a quarter of the left hand side (orange area) is gated with $\mu=0.1$. (bottom) The current flowing along ${\hat{\bf x}}$ direction at the $y=1$ edge $\langle J_{x}^{0}(y=1)\rangle$ as a function of $x$. (b) The same as (a), but at bulk gap $M=-0.2$ that yields a longer edge state decay length $\xi$. } 
\label{fig:nonlocal_edge_current_results}
\end{center}
\end{figure}

An experimentally relevant system that belongs to 2D class A is the thin film of magnetic TIs of few quintuple layers that manifests quantum anomalous Hall effect (QAHE)\cite{Chang13,Liu16,He18}. Although the equilibrium edge current will not show up in the nonequilibrium transport measurements, the magnetic field it produces should in principle be measurable. Assuming a Chern insulator quantum dot of dimension $N_{x}a\sim 100$nm can be realized by these systems\cite{Ferreira13}, the edge current in the quantum dot is quantized as a function of gate voltage $\mu$ similar to that shown in Fig.~\ref{fig:Chern_results} (c). A typical Fermi velocity $v_{F}\sim 10^{5}$m/s gives the edge current quantum $\Delta J_{x}^{0}\sim 10^{-7}$A. Ampere's law then gives an increase of magnetic field at the center $\Delta B=\mu_{0}\Delta J_{x}^{0}/2\pi N_{x}a\sim 10^{-7}$T, and consequently an increase of magnetic flux $\Phi_{B}\sim \pi (N_{x}a)^{2}B\sim 10^{-13}$Wb$\sim 100\Phi_{B}^{0}$ per edge current quantum, where $\Phi_{B}^{0}$ is the magnetic flux quantum. Thus the magnetic flux through the quantum dot as a function of gate voltage is also quantized, which should be measurable by magnetic flux detectors such as the superconducting quantum interference device (SQUID), as indicated schematically in Fig.~\ref{fig:Chern_results} (d).

\subsection{Nonlocal edge current in partially gated topological quantum dot}

We now examine a Chern insulator quantum dot that is partially gated, and show that the edge current created in the gated area can extend into the ungated area over length scale $\xi$. To elaborate this statement, in Fig.~\ref{fig:nonlocal_edge_current_results} we show the local current pattern in a $N_{x}\times N_{y}=32\times 8$ quantum dot with OBC imposed in both ${\hat{\bf x}}$ and ${\hat{\bf y}}$ directions. A quarter of the area is gated to have $\mu=0.1$, whereas the ungated area is kept at $\mu=0$. The result indicates that the boundary between the gated and ungated areas serves as an edge, and an edge current is created in both sides of the edge even though only one side has a finite chemical potential $\mu$. The circulating pattern of the current decays into the ungated area over length $\xi$, as concluded from comparing the $M=-1$ (small $\xi\approx a$) and $M=-0.2$ (large $\xi\approx 5a$) cases that show different decay lengths. The current flowing along ${\hat{\bf x}}$ direction at the $y=1$ edge $\langle J_{x}^{0}(y=1)\rangle$ also confirms this decaying behavior, as shown in Fig.~\ref{fig:nonlocal_edge_current_results}. Thus we anticipate that this partially gated TI quantum dot can serve as a nonlocal edge current generator, with the range of the edge current controllable by the insulating gap $M$. Finally, we remark that in all cases, the continuity equation in Eq.~(\ref{ndot_commutator}) is satisfied $\dot{n}_{i}=0$ at any site $i$, provided that the current flowing along positive $J_{i,i+\delta}^{0}$ and negative bonds $J_{i,i-\delta}^{0}$ are both taken into account.

\begin{figure}[ht]
\begin{center}
\includegraphics[clip=true,width=0.99\columnwidth]{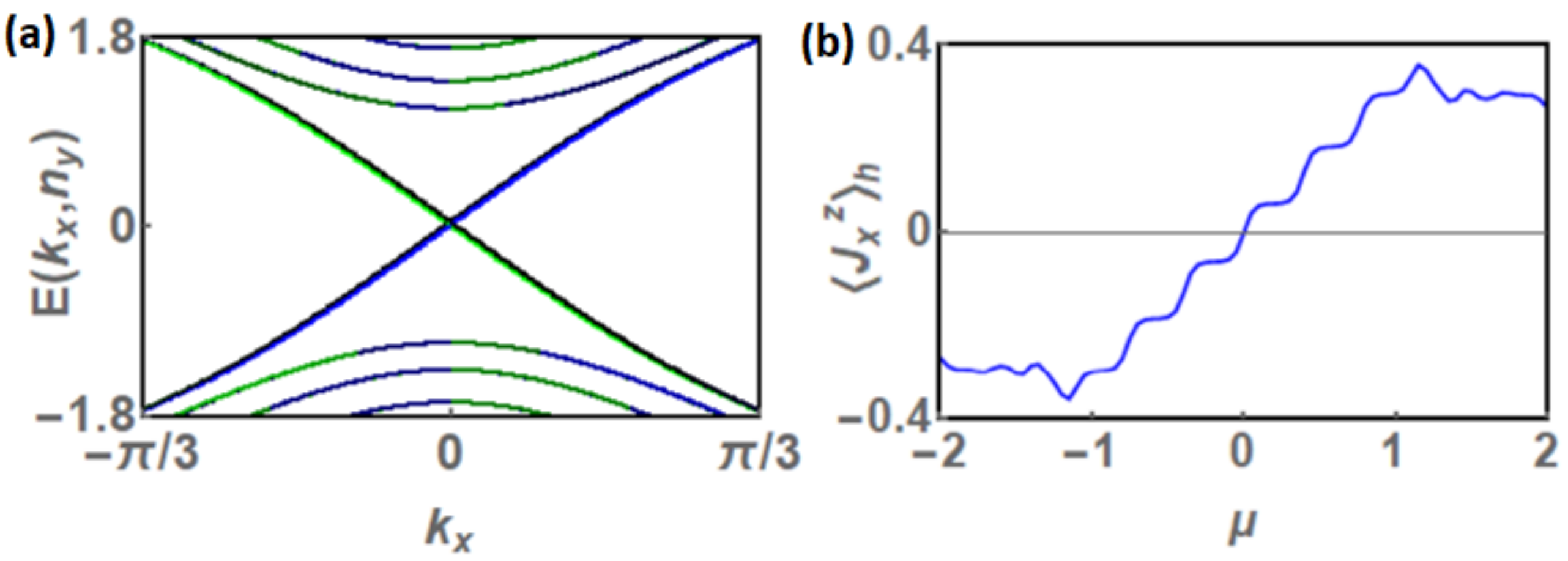}
\caption{(a) The band structure of a strip of BHZ model, with blue and green colors indicating the weight of spin up and down for the wave function closer the $y=1$ edge. (b) Quantization of edge spin current in the BHZ strip as a function of the chemical potential, which is essentially the same as Fig.~\ref{fig:Chern_results} (c) up to an overall prefactor.  } 
\label{fig:BHZ_results}
\end{center}
\end{figure}

\subsection{2D class AII}

For 2D class AII, we use the prototype Bernevig-Hughes-Zhang (BHZ) model as an example\cite{Bernevig06}, which has basis $\psi=\left(s\uparrow,p\uparrow,s\downarrow,p\downarrow\right)^{T}$ and the Dirac matrices 
\begin{eqnarray}
\Gamma^{\ell}=\left\{\sigma^{z}\otimes s^{x},I\otimes s^{y},I\otimes s^{z},\sigma^{x}\otimes s^{x},\sigma^{y}\otimes s^{x}\right\}.
\end{eqnarray}
where $\sigma^{b}$ and $s^{b}$ are Pauli matrices in the spin and orbital spaces, respectively. The continuum model reads 
\begin{eqnarray}
H({\bf k})&=&A\sin k_{x}\Gamma^{1}+A\sin k_{y}\Gamma^{2}
\nonumber \\
&+&\left(M-4B+2B\cos k_{x}+2B\cos k_{y}\right)\Gamma^{3}\;,
\label{BHZ_Dirac_Hamiltonian}
\end{eqnarray}
which regularizes to give the square lattice model
\begin{eqnarray}
&&H=\sum_{i\sigma}t\left\{-i\sigma c_{is\sigma}^{\dag}c_{i+ap\sigma}
-i\sigma c_{ip\sigma}^{\dag}c_{i+as\sigma}+h.c.\right\}
\nonumber \\
&&+\sum_{i\sigma}t\left\{-c_{is\sigma}^{\dag}c_{i+bp\sigma}+c_{ip\sigma}^{\dag}c_{i+bs\sigma}+h.c.\right\}
\nonumber \\
&&+\sum_{i\sigma}\left(M+4t'-\mu\right)c_{is\sigma}^{\dag}c_{is\sigma}
+\sum_{i}\left(-M-4t'-\mu\right)c_{ip\sigma}^{\dag}c_{ip\sigma}
\nonumber \\
&&-\sum_{i\sigma\delta}t'\left\{c_{is\sigma}^{\dag}c_{i+\delta s\sigma}-c_{ip\sigma}^{\dag}c_{i+\delta p\sigma}+h.c.\right\},
\label{Hamiltonian_2DclassAII}
\end{eqnarray}
where $t=A/2$ and $t'=-B$. Here $I=\left\{s,p\right\}$ is the orbital index, $\delta=\left\{a,b\right\}$ denotes the lattice constant along the two planar directions, $\sigma=\left\{\uparrow,\downarrow\right\}=\left\{+,-\right\}$ is the spin index, $i=\left\{x,y\right\}$ denotes the planar position. We use the parameters $-t=t'=-M=1$ and temperature $k_{B}T=0.03$. Once again the chemical potential $\mu$ will be crucial for the existence of edge spin current.

The band structure at $N_{y}=8$ is shown in Fig.~\ref{fig:BHZ_results}. The edge states consist of counter propagating spins polarized along ${\hat{\bf z}}$, as can be seen from the expectation value of $\sigma^{z}$ of $|k_{x},n_{y}\rangle$ near the $y=1$ edge
\begin{eqnarray}
\tilde{m}_{k_{x},n_{y}}^{z}=\sum_{1\leq y\leq N_{y}/2}\sigma_{k_{x},n_{y}}^{z}.
\end{eqnarray}
The color in the band structure of Fig.~\ref{fig:BHZ_results} (b) indicates the spin up (blue) and down (green) of $\tilde{m}_{k_{x},n_{y}}^{z}$. The branch of Dirac cone of positive group velocity has spin up and that of negative group velocity has spin down, as expected (note that the Dirac cone of black color is that localized at the other edge $y=N_{y}$). Besides, the topmost valence band at $k_{x}>0$ with negative group velocity has more spin up component, whereas that at $k_{x}<0$ with positive group velocity have more spin down component. This indicates at least some valence bands contribute to a spin current {\it against} that contributed from the edge states. We further quantify the two contributions by considering the local spin current operators constructed from the equation of motion of the spin density $m_{i}^{a}=\sum_{I}c_{iI\alpha}^{\dag}\sigma_{\alpha\beta}^{a}c_{iI\beta}$
\begin{eqnarray}
\dot{m}_{i}^{a}=\frac{i}{\hbar}\left[H,m_{i}^{a}\right]=-{\boldsymbol\nabla}\cdot{\bf J}_{i}^{a}
=-\frac{1}{a}\sum_{\delta}\left(J_{i,i+\delta}^{a}+J_{i,i-\delta}^{a}\right).
\nonumber \\
\label{Mdot_commutator}
\end{eqnarray}
Because the edge states are spin polarized along ${\hat{\bf z}}$, we investigate the local spin current operators $J_{x}^{z}\equiv J_{i,i+a}^{z}$ and $J_{y}^{z}\equiv J_{i,i+b}^{z}$ that are given by
\begin{eqnarray}
&&J_{x}^{z}=\frac{a}{\hbar}\sum_{\sigma}t\left\{c_{is\sigma}^{\dag}c_{i+ap\sigma}
+c_{ip\sigma}^{\dag}c_{i+as\sigma}+h.c.\right\}
\nonumber \\
&&+\frac{a}{\hbar}\sum_{\sigma}t'\left\{-i\sigma c_{is\sigma}^{\dag}c_{i+as\sigma}
+i\sigma c_{ip\sigma}^{\dag}c_{i+ap\sigma}+h.c.\right\},
\nonumber \\
&&J_{y}^{z}=\frac{b}{\hbar}\sum_{\sigma}t\left\{-i\sigma c_{is\sigma}^{\dag}c_{i+bp\sigma}+i\sigma c_{ip\sigma}^{\dag}c_{i+bs\sigma}+h.c.\right\}
\nonumber \\
&&+\frac{b}{\hbar}\sum_{\sigma}t'\left\{-i\sigma c_{is\sigma}^{\dag}c_{i+bs\sigma}
+i\sigma c_{ip\sigma}^{\dag}c_{i+bp\sigma}+h.c.\right\}.
\end{eqnarray}
The spatial profile of the edge spin current evaluated with an energy cutoff $E_{cut}$, as in Eq.~(\ref{current_expectation_Ecut}), is essentially the same as that in Fig.~\ref{fig:Chern_results} (b) for the Chern insulators, up to an overall prefactor. This is not surprising, since the BHZ model is basically two copies (spin up and down) of the Chern insulator, and we also see that the contribution from the valence bands cancels out exactly that from the edge states, rendering no edge spin current if the Dirac cone resides at the chemical potential $\mu=0$. The quantized edge spin current $\langle J_{x}^{z}\rangle_{h}=\sum_{1\leq y\leq N_{y}/2}\langle J_{x}^{z}\rangle$ as a function of chemical potential $\mu$ is shown in Fig.~\ref{fig:BHZ_results} (b), which has the same number of plateaux and edge spin current quantum as that in Eqs.~(\ref{number_of_plateaux}) and (\ref{height_of_plateaux}), except one replaces the electron charge by the Bohr magneton $e\rightarrow \mu_{B}$ in these equations.

The equilibrium edge spin current $J_{x}^{z}$ does not show up in nonequilibrium transport or spin transport measurements, but the electric dipolar field\cite{Hirsch90,Hirsch99_2,Meier03,Schutz03,Sun04,Chen14_inductance_spin_current} it produces should in principle be measurable. This can be understood by considering that a magnetic moment ${\bf m}$ moving with velocity ${\bf v}$ produces an electric dipole moment in the rest frame
\begin{eqnarray}
{\bf p}=\frac{\gamma}{c^{2}}{\bf v}\times{\bf m},
\label{electric_dipole}
\end{eqnarray}
where $\gamma=(1-v^{2}/c^{2})^{-1/2}$, and subsequently a dipolar field at position ${\bf r}$ from the moving magnetic moment
\begin{eqnarray}
{\bf E_{m}(r)}=\frac{3{\hat{\bf r}}({\bf p}\cdot{\hat{\bf r}})-{\bf p}}{4\pi\epsilon_{0}r^{3}}.
\label{electric_dipolar_field}
\end{eqnarray}
Approximating the edge along ${\hat{\bf x}}$ as a long straight wire, the edge spin current quantum $\Delta J_{x}^{z}=\mu_{B}v_{F}/N_{x}a$ corresponds to an electric dipolar field\cite{Meier03}
\begin{eqnarray}
\Delta{\bf E}_{m}({\bf r})=\frac{\mu_{0}\Delta J_{x}^{z}}{2\pi r^{2}}(0,\cos 2\phi,-\sin 2\phi),
\end{eqnarray}
where $\sin\phi=y/r$, $\cos\phi=z/r$, and $r=\sqrt{y^{2}+z^{2}}$. Assuming a quantum dot of size $N_{x}\sim 10^{2}$ and a typical Fermi velocity $v_{F}\sim 10^{5}$m/s, the electric dipolar field at a distance $r\sim$ nm away from the edge is $|\Delta{\bf E}_{m}|\sim 1$V/m. Thus two points that are $\left\{{\bf r}_{1},{\bf r}_{2}\right\}\sim$ nm away from the edge and are ${\bf r}_{1}-{\bf r}_{2}\sim$ nm apart would experience a voltage drop $\Delta V\sim $ nV, which should be measurable by local high impedance voltmeters, assuming the gate voltage does not affect the measurement.

The local spin current promoted by impurities and edge confining potential follows that discussed in Sec.~\ref{sec:impurity_confining}. The nonlocal edge spin current in partially gated BHZ quantum dot is also given by that of the Chern insulator in Fig.~\ref{fig:nonlocal_edge_current_results}, and the extension of the edge spin current into the ungated area is again characterized by the decay length $\xi$. Note that the BHZ model at parameters $M\sim -0.01$eV and $\hbar v_{F}/a\sim$ eV is relevant to the HgTe quantum well\cite{Konig07,Konig08} or recently proposed III-V semiconductor quantum well\cite{Candido18} at few nanometers thickness, which yields a decay length of the order of $\xi\sim 100a\sim 100$nm. Thus if a partially gated HgTe quantum well of area $\sim 100$nm$\times 100$nm can be fabricated, it can serve as a dissipationless nonlocal spin current generator of $\sim 100$nm range, which is fairly significant. In contrast, the recently proposed monolayer WTe$_{2}$ has $M\sim -0.1$eV and $\hbar v_{F}/a\sim 0.1$eV\cite{Qian14,Zheng16,Tang17}, so $\xi\sim a$ is of the order of lattice constant, which may not be the ideal material for the nonlocal spin current generator.

%{\cblue (1) Invoke my inductance of spin current paper. Can we measure the electric flux vector produced by the spin current? }

\subsection{3D class AII} 

\begin{figure}[ht]
\begin{center}
\includegraphics[clip=true,width=0.99\columnwidth]{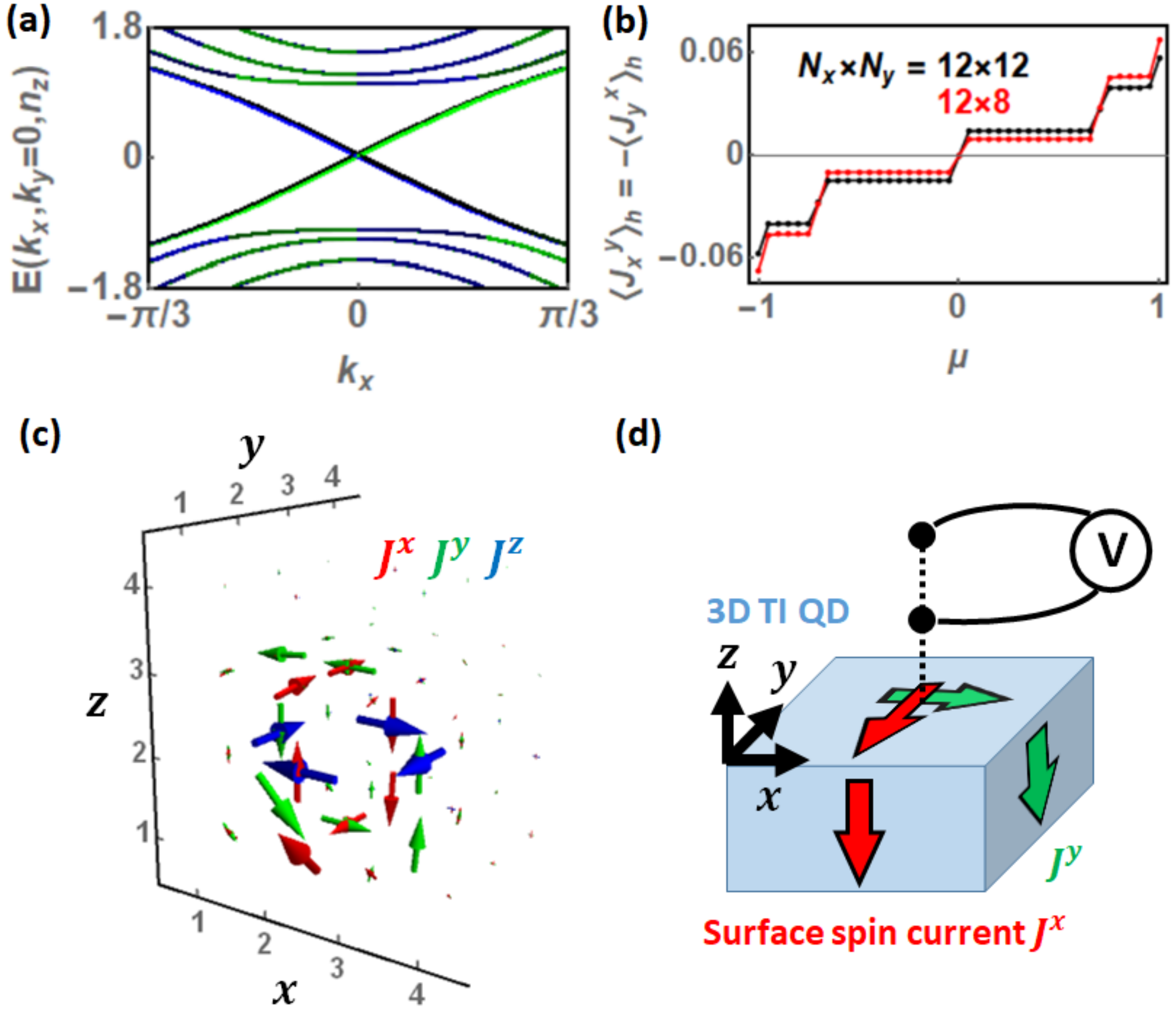}
\caption{(a) The band structure of 3D TI along $k_{x}$ at $k_{y}=0$, with blue and green colors indicating the sign of spin polarization along ${\hat{\bf y}}$ closer to the top surface $z=1$. (b) The spin current $\langle J_{x}^{y}\rangle_{h}=-\langle J_{y}^{x}\rangle_{h}$ in the half-space $1\leq z\leq N_{z}/2$ as a function of chemical potential $\mu$ in two different lattice sizes. (c) The 3D vortex of spin current $\left\{J^{x},J^{y},J^{z}\right\}$ produced around a point-like impurity in a 3D TI. (d) The proposal of measuring the electric dipolar field produced by the surface spin current, where two points that are $\sim$ nm above the surface of the quantum dot of dimension $L\sim 100$nm has a voltage difference of $\sim$ nV. } 
\label{fig:3DTI_results}
\end{center}
\end{figure}

We next consider 3D TIs in class AII, which are relevant to materials such as Bi$_{2}$Se$_{3}$ and Bi$_{2}$Te$_{3}$. These materials as made have a single edge state Dirac cone with the Dirac point located $\sim 0.1$eV away from the chemical potential\cite{Chen09,Analytis10,Zhang10,Pan11}, but the position of Dirac point can be efficiently engineered by doping\cite{Hsieh09,Zhang11,Kondou16}. We will consider the theoretical model for the low energy sector described by the $\Gamma$-matrices\cite{Zhang09,Liu10}
\begin{eqnarray}
\Gamma^{\ell}=\left\{\sigma^{x}\otimes\tau^{x},\sigma^{y}\otimes\tau^{x},\sigma^{z}\otimes\tau^{x},
I_{\sigma}\otimes\tau^{y},I_{\sigma}\otimes\tau^{z}\right\},
\nonumber \\
\end{eqnarray} 
where the spinor is written as $\psi_{\bf k}=\left(c_{{\bf k}P1_{-}^{+}\uparrow},c_{{\bf k}P2_{+}^{-}\uparrow},c_{{\bf k}P1_{-}^{+}\downarrow},c_{{\bf k}P2_{+}^{-}\downarrow}\right)^{T}\equiv\left(c_{{\bf k}s\uparrow},c_{{\bf k}p\uparrow},c_{{\bf k}s\downarrow},c_{{\bf k}p\downarrow}\right)^{T}$, where $s$ and $p$ are abbreviations for the $P1_{-}^{+}$ and $P2_{+}^{-}$ orbitals in real materials. The low energy Hamiltonian given by ${\bf k\cdot p}$ theory is
\begin{eqnarray}
\hat{H}&=&\left(M+M_{1}k_{z}^{2}+M_{2}k_{x}^{2}+M_{2}k_{y}^{2}\right)\Gamma^{5}
\nonumber \\
&+&B_{0}\Gamma^{4}k_{z}
+A_{0}\left(\Gamma^{1}k_{y}-\Gamma^{2}k_{x}\right),
\label{3D_TI_H0_H1}
\end{eqnarray}
where we keep only lowest order terms. The regularization on a cubic lattice yields
\begin{eqnarray}
&&H=-\sum_{iI\sigma}\mu c_{iI\sigma}^{\dag}c_{iI\sigma}+\sum_{i\in TI,\sigma}\tilde{M}\left\{c_{is\sigma}^{\dag}c_{is\sigma}-c_{ip\sigma}^{\dag}c_{ip\sigma}\right\}
\nonumber \\
&&+\sum_{i\in TI,I}t_{\parallel}\left\{c_{iI\uparrow}^{\dag}c_{i+a\overline{I}\downarrow}
-c_{i+aI\uparrow}^{\dag}c_{i\overline{I}\downarrow}+h.c.\right\}
\nonumber \\
&&+\sum_{i\in TI,I}t_{\parallel}\left\{-ic_{iI\uparrow}^{\dag}c_{i+b\overline{I}\downarrow}
+ic_{i+bI\uparrow}^{\dag}c_{i\overline{I}\downarrow}+h.c.\right\}
\nonumber \\
&&+\sum_{i\in TI,\sigma}t_{\perp}\left\{-c_{is\sigma}^{\dag}c_{i+cp\sigma}+c_{i+cs\sigma}^{\dag}c_{ip\sigma}+h.c.\right\}
\nonumber \\
&&-\sum_{i\in TI,\sigma}M_{1}\left\{c_{is\sigma}^{\dag}c_{i+cs\sigma}-c_{ip\sigma}^{\dag}c_{i+cp\sigma}+h.c.\right\}
\nonumber \\
&&-\sum_{i\in TI,\delta,\sigma}M_{2}\left\{c_{is\sigma}^{\dag}c_{i+\delta s\sigma}-c_{ip\sigma}^{\dag}c_{i+\delta p\sigma}+h.c.\right\},
\label{3DTIFMM_Hamiltonian}
\end{eqnarray}
where $\tilde{M}=M+2M_{1}+4M_{2}$, $t_{\parallel}=A_{0}/2$, $t_{\perp}=B_{0}/2$, $I=\left\{s,p\right\}$ and $\overline{I}=\left\{p,s\right\}$ are the orbital indices, $\delta=\left\{a,b,c\right\}$ denotes the lattice constants, $\sigma=\left\{\uparrow,\downarrow\right\}$ is the spin index, and $\mu$ is the chemical potential crucial for the existence of the surface spin current. We use the band parameters $t_{\parallel}=t_{\perp}=M_{1}=M_{2}=-M=1$ such that conclusions can be drawn by numerics done in a accessible lattice size of the order of $N_{x}\times N_{y}\times N_{z}\sim 10\times 10\times 10$, but we emphasize that the conclusions are robust against changing the parameters within the same order of magnitude. Choosing OBC along ${\hat{\bf z}}$ and PBC along ${\hat{\bf x}}$ and ${\hat{\bf y}}$, the surface state of momentum ${\bf k}_{\parallel}=(k_{x},k_{y})$ is spin polarized along ${\hat{\bf z}}\times{\hat{\bf k}}_{\parallel}$, which speculates a spin current $J_{x}^{y}$ flowing along ${\hat{\bf x}}$ and polarized along ${\hat{\bf y}}$, and a $J_{y}^{x}$ flowing along ${\hat{\bf y}}$ and polarized along ${\hat{\bf x}}$ of equal magnitude. In Fig.~\ref{fig:3DTI_results} (a), we use the band structure along $k_{x}$ at $k_{y}=0$ to elaborate the spin-momentum locking, where the sign of the spin polarization along ${\hat{\bf y}}={\hat{\bf z}}\times{\hat{\bf x}}$ close to the surface at $z=1$
\begin{eqnarray}
\tilde{m}_{k_{x},k_{y}=0,n_{z}}^{y}=\sum_{1\leq z\leq N_{z}/2}\sigma_{k_{x},k_{y}=0,n_{z}}^{y}.
\end{eqnarray}
is indicated by blue (positive) and green (negative). Once again the spin polarization of the Dirac cone is evident, and it is also clear that some valence bands have the same spin polarization as the Dirac cone but opposite group velocities, hence producing a spin current against that produced by the surface states.

To quantify the effect of valence bands, we turn to the spin current operators constructed from Eq.~(\ref{Mdot_commutator}). The spin currents flowing along the positive bonds $J_{x}^{y}\equiv J_{i,i+a}^{y}$ and $J_{y}^{x}\equiv J_{i,i+b}^{x}$ are given by
\begin{eqnarray}
&&J_{x}^{y}=\frac{ia}{\hbar}\left\{it_{\parallel}\sum_{I}\left[
c_{iI\uparrow}^{\dag}c_{i+a\overline{I}\uparrow}
+c_{iI\downarrow}^{\dag}c_{i+a\overline{I}\downarrow}\right]-h.c.\right\}
\nonumber \\
&&+\frac{ia}{\hbar}\left\{iM_{2}\sum_{\sigma}\left[
\sigma c_{is\sigma}^{\dag}c_{i+as\overline{\sigma}}-\sigma c_{ip\sigma}^{\dag}c_{i+ap\overline{\sigma}}\right]-h.c.\right\}
\nonumber \\
&&J_{y}^{x}=\frac{ib}{\hbar}\left\{it_{\parallel}\sum_{I}\left[
-c_{iI\uparrow}^{\dag}c_{i+b\overline{I}\uparrow}
-c_{iI\downarrow}^{\dag}c_{i+b\overline{I}\downarrow}\right]-h.c.\right\}
\nonumber \\
&&+\frac{ib}{\hbar}\left\{M_{2}\sum_{\sigma}\left[
-c_{is\sigma}^{\dag}c_{i+bs\overline{\sigma}}
+c_{ip\sigma}^{\dag}c_{i+bp\overline{\sigma}}\right]-h.c.\right\}.\;\;\;\;
\end{eqnarray}
where $I=\left\{s,p\right\}$ and $\overline{I}=\left\{p,s\right\}$. Using Eq.~(\ref{current_expectation_Ecut}), we obtain a finite spin current as $E_{cut}<|M|$ that includes only the surface states, and eventually vanishes at large $E_{cut}$ that includes all the bands. The spin current is again finite when the chemical potential $\mu$ is nonzero, and changes sign when $\mu$ sweeps through the Dirac point, as shown in Fig.~\ref{fig:3DTI_results} (b). Thus we also expect that several factors in realistic 3D TIs, such as impurities\cite{Wray11}, surface band bending\cite{Bahramy12}, and local electron and hole puddles\cite{Beidenkopf11}, helps to promote the surface spin current, since they shift Dirac point away from the chemical potential globally or locally. As an example, the 3D vortex of the spin currents formed around a single impurity is shown in Fig.~\ref{fig:3DTI_results} (c), where one sees that the spin current $J^{a}$ that is polarized along ${\hat{\bf a}}$ wraps around the ${\hat{\bf a}}$ axis. We should also emphasize that despite the absence of the equilibrium edge spin current at $\mu=0$, the edge states can still perform nonequilibrium transport and spintronic effects, such as the conductance due to a bias voltage\cite{Zhou17} and the current-induced spin accumulation\cite{Chen20_TI_Edelstein,Zegarra20} at $\mu=0$. This is because these nonequlibrium spintronic effects only involve the edge states near the chemical potential due to the derivative of the Fermi function, and hence the valence bands are irrelevant.

The quantization of edge spin current $\langle J_{x}^{y}\rangle_{h}=\sum_{1\leq z\leq N_{z}/2}\langle J_{x}^{y}\rangle=-\langle J_{y}^{x}\rangle_{h}$ as a function of $\mu$ is again evident, as shown in Fig.~\ref{fig:3DTI_results} (b). The number of plateaux $N_{p}$ and the surface spin current quantum $\Delta J_{x}^{y}$ are irregular functions of $\mu$ (the steps are not evenly spaced), and changes with the shape of the quantum dot, as can be seen by comparing the case of $N_{x}\times N_{y}\times N_{z}=12\times 12\times 8$ with that of $12\times 8\times 8$. This is presumably due to the irregularity in the number of states that can fit into the Dirac cone, which is best described in a cylindrical coordinate, but the lattice is cubic. Nevertheless, the trend of more plateaux and smaller surface spin current quantum at larger quantum dot, similar to the 2D cases described by Eqs.~(\ref{number_of_plateaux}) and (\ref{height_of_plateaux}), is qualitatively correct.

The surface spin current also produces an electric dipolar field according to Eqs.~(\ref{electric_dipole}) and (\ref{electric_dipolar_field}). Consider a cubic quantum dot of 3D TI whose top surface area $L^{2}$ is confined in the region $-L/2\leq x\leq L/2$ and $-L/2\leq y\leq L/2$. A uniform $\langle J_{x}^{y}\rangle$ (e.g., produced by doping) flowing in the top surface can be regarded as a current $I_{m}=\mu_{B}\langle J_{x}^{y}\rangle/a=m\rho v$ of magnetic moment ${\bf m}$ of planar density $\rho$ polarized in ${\hat{\bf y}}$ and moving along ${\hat{\bf x}}$. A point located at $z$ above the center of the quantum dot, as shown schematically in Fig.~\ref{fig:3DTI_results} (d), experiences an electric field along ${\hat{\bf z}}$ direction
\begin{eqnarray}
E^{z}(z)&=&\frac{\mu_{0}I_{m}}{4\pi}\left\{\frac{8\sqrt{2}L^{2}}{\sqrt{L^{2}+2z^{2}}(L^{2}+4z^{2})}\right.
\nonumber \\
&&+\frac{4}{z}{\rm arccot}\left(\frac{2\sqrt{2}z\sqrt{L^{2}+2z^{2}}}{L^{2}}\right)
\nonumber \\
&&+\left.-\frac{4}{z}{\rm arctan}\left(\frac{L^{2}}{2\sqrt{2}z\sqrt{L^{2}+2z^{2}}}\right)\right\}.
\label{dipolar_field_of_3DTI_QD}
\end{eqnarray}
In the limit $z\ll L$, i.e., very close to the top surface of a large quantum dot, one has $E^{z}\approx \mu_{0}I_{m}/L$. Given that $\langle J_{x}^{y}\rangle$ in Fig.~\ref{fig:3DTI_results} (b) is of the order of $0.01t/\hbar$, the corresponding magnetization current is $I_{m}\sim 0.1$Cm/s$^{2}$. For a quantum dot of dimension $L\sim 100a\sim 10^{-7}$m, the magnetization current produces an electric field $E^{z}\approx \mu_{0}I_{m}/L\sim 1$V/m near the top surface. Thus for two points $\left\{z_{1},z_{2}\right\}$ that are nm away from the surface and are nm apart, as that in in Fig.~\ref{fig:3DTI_results} (d), the voltage difference between them is $\Delta V\sim$nV, which is readily measurable. Interestingly, Eq.~(\ref{dipolar_field_of_3DTI_QD}) also implies that the surface spin current of an infinitely large TI $L\rightarrow\infty$ does not produce a electric dipolar field $E^{z}\rightarrow 0$. So the aforementioned voltage difference can only be observed in TIs of finite size.

\subsection{Implications for real TIs}

The band structures resulted from regularizing the low energy Dirac Hamiltonian to the entire BZ, as shown in Fig.~\ref{fig:Chern_results} (a), \ref{fig:BHZ_results} (a), and \ref{fig:3DTI_results} (a) certainly cannot capture the band structures in real TIs. Thus our conclusion of the exact cancellation of the edge current by valence bands is valid only within these regularized lattice models. In real TIs, we suspect that the valence bands can contribute to an edge current flowing in either direction depending on details of the material. To settle this issue, a feasible approach would be that one starts from the realistic tight-binding models constructed from certain {\it ab initio} calculations for specific materials\cite{Zhang09,Liu10}, extract the current operators from Eqs.~(\ref{ndot_commutator}) and (\ref{Mdot_commutator}), then calculate the equilibrium edge currents from Eq.~(\ref{current_expectation_Ecut}). How the edge current depends on the chemical potential $\mu$ can be easily verified, and the effect of inhomogeneity can be simulated by adding appropriate impurities or edge confining potentials into these realistic lattice models. Although it is not our purpose to perform this kind of calculation in the present work, the important point drawn from our simplified approach is that one cannot judge the amount and direction of the edge currents solely from the edge state Dirac cone, and the valence bands generally contribute to all equilibrium properties of TIs, even for those at the edge or surface.

\section{Conclusions} 

In summary, we demonstrate that although the edge states circulating the boundary naturally speculate the existence of an equilibrium edge current or edge spin current\cite{Buttiker09,Sonin11,Ando13,Maekawa17}, in reality the edge current is not solely determined by the edge states. We point out that the valence bands also contribute significantly to the edge current. If the low energy Dirac cone is regularized into a lattice model in a straightforward manner, as has been done in many theoretical models in different symmetry classes, then the contribution from the valence bands exactly cancels out that from the edge states, rendering no edge current. This statement is valid regardless the temperature and band parameters of the regularized lattice model. Our result therefore serves as a warning that for any equilibrium property of TIs, the contribution from the valence bands should not be ignored, since equilibrium properties are not solely determined by the edge states.

Despite the regularized lattice models are not expected to capture the band structures in real TIs, the following features we reveal are anticipated to manifest in experiments. %Firstly, the cancellation from the valence bands suggests that there exists a specific value $\mu_{c}$ of the chemical potential at which the edge current or edge spin current vanishes, which may not be exactly at the Dirac point in reality due to complications from the band structures. The edge current changes the direction of flow as chemical potential sweeps through $\mu_{c}$. 
Firstly, the global or local variation of the chemical potential, such as that caused by gating, doping, disorder, surface band bending, or edge confining potential, all can promote the edge current globally or locally. Particularly in gated TI quantum dots, the edge current as a function of the gate voltage is geometrically quantized, with the number of plateaux and edge current quantum determined by the size of the quantum dot, Fermi velocity, and the insulating gap. Moreover, the edge current in partially gated quantum dots can extend into the ungated region over the decay length of the edge state, hence acting as electrically controllable dissipationless current or spin current generators that may have applications in mesoscopic electronic or spintronic devices.

\section{Acknowledgement}

The author acknowledges stimulating discussions with J. C. Egues and A. Zegarra, and the financial support from the productivity in research fellowship from CNPq.

%{\cblue (1) Further investigations: other types of lattice model such as those based on honeycomb lattice. }

\bibliography{Literatur}

%merlin.mbs apsrev4-1.bst 2010-07-25 4.21a (PWD, AO, DPC) hacked
%Control: key (0)
%Control: author (8) initials jnrlst
%Control: editor formatted (1) identically to author
%Control: production of article title (-1) disabled
%Control: page (0) single
%Control: year (1) truncated
%Control: production of eprint (0) enabled
\begin{thebibliography}{59}%
\makeatletter
\providecommand \@ifxundefined [1]{%
 \@ifx{#1\undefined}
}%
\providecommand \@ifnum [1]{%
 \ifnum #1\expandafter \@firstoftwo
 \else \expandafter \@secondoftwo
 \fi
}%
\providecommand \@ifx [1]{%
 \ifx #1\expandafter \@firstoftwo
 \else \expandafter \@secondoftwo
 \fi
}%
\providecommand \natexlab [1]{#1}%
\providecommand \enquote  [1]{``#1''}%
\providecommand \bibnamefont  [1]{#1}%
\providecommand \bibfnamefont [1]{#1}%
\providecommand \citenamefont [1]{#1}%
\providecommand \href@noop [0]{\@secondoftwo}%
\providecommand \href [0]{\begingroup \@sanitize@url \@href}%
\providecommand \@href[1]{\@@startlink{#1}\@@href}%
\providecommand \@@href[1]{\endgroup#1\@@endlink}%
\providecommand \@sanitize@url [0]{\catcode `\\12\catcode `\$12\catcode
  `\&12\catcode `\#12\catcode `\^12\catcode `\_12\catcode `\%12\relax}%
\providecommand \@@startlink[1]{}%
\providecommand \@@endlink[0]{}%
\providecommand \url  [0]{\begingroup\@sanitize@url \@url }%
\providecommand \@url [1]{\endgroup\@href {#1}{\urlprefix }}%
\providecommand \urlprefix  [0]{URL }%
\providecommand \Eprint [0]{\href }%
\providecommand \doibase [0]{http://dx.doi.org/}%
\providecommand \selectlanguage [0]{\@gobble}%
\providecommand \bibinfo  [0]{\@secondoftwo}%
\providecommand \bibfield  [0]{\@secondoftwo}%
\providecommand \translation [1]{[#1]}%
\providecommand \BibitemOpen [0]{}%
\providecommand \bibitemStop [0]{}%
\providecommand \bibitemNoStop [0]{.\EOS\space}%
\providecommand \EOS [0]{\spacefactor3000\relax}%
\providecommand \BibitemShut  [1]{\csname bibitem#1\endcsname}%
\let\auto@bib@innerbib\@empty
%</preamble>
\bibitem [{\citenamefont {Hasan}\ and\ \citenamefont {Kane}(2010)}]{Hasan10}%
  \BibitemOpen
  \bibfield  {author} {\bibinfo {author} {\bibfnamefont {M.~Z.}\ \bibnamefont
  {Hasan}}\ and\ \bibinfo {author} {\bibfnamefont {C.~L.}\ \bibnamefont
  {Kane}},\ }\href {\doibase 10.1103/RevModPhys.82.3045} {\bibfield  {journal}
  {\bibinfo  {journal} {Rev. Mod. Phys.}\ }\textbf {\bibinfo {volume} {82}},\
  \bibinfo {pages} {3045} (\bibinfo {year} {2010})}\BibitemShut {NoStop}%
\bibitem [{\citenamefont {Qi}\ and\ \citenamefont {Zhang}(2011)}]{Qi11}%
  \BibitemOpen
  \bibfield  {author} {\bibinfo {author} {\bibfnamefont {X.-L.}\ \bibnamefont
  {Qi}}\ and\ \bibinfo {author} {\bibfnamefont {S.-C.}\ \bibnamefont {Zhang}},\
  }\href {\doibase 10.1103/RevModPhys.83.1057} {\bibfield  {journal} {\bibinfo
  {journal} {Rev. Mod. Phys.}\ }\textbf {\bibinfo {volume} {83}},\ \bibinfo
  {pages} {1057} (\bibinfo {year} {2011})}\BibitemShut {NoStop}%
\bibitem [{\citenamefont {Bernevig}\ and\ \citenamefont
  {Hughes}(2013)}]{Bernevig13}%
  \BibitemOpen
  \bibfield  {author} {\bibinfo {author} {\bibfnamefont {B.~A.}\ \bibnamefont
  {Bernevig}}\ and\ \bibinfo {author} {\bibfnamefont {T.~L.}\ \bibnamefont
  {Hughes}},\ }\href@noop {} {\emph {\bibinfo {title} {Topological Insulators
  and Topological Superconductors}}}\ (\bibinfo  {publisher} {Princeton
  University Press},\ \bibinfo {year} {2013})\BibitemShut {NoStop}%
\bibitem [{\citenamefont {Jackiw}\ and\ \citenamefont
  {Rebbi}(1976)}]{Jackiw76}%
  \BibitemOpen
  \bibfield  {author} {\bibinfo {author} {\bibfnamefont {R.}~\bibnamefont
  {Jackiw}}\ and\ \bibinfo {author} {\bibfnamefont {C.}~\bibnamefont {Rebbi}},\
  }\href {\doibase 10.1103/PhysRevD.13.3398} {\bibfield  {journal} {\bibinfo
  {journal} {Phys. Rev. D}\ }\textbf {\bibinfo {volume} {13}},\ \bibinfo
  {pages} {3398} (\bibinfo {year} {1976})}\BibitemShut {NoStop}%
\bibitem [{\citenamefont {Schnyder}\ \emph {et~al.}(2008)\citenamefont
  {Schnyder}, \citenamefont {Ryu}, \citenamefont {Furusaki},\ and\
  \citenamefont {Ludwig}}]{Schnyder08}%
  \BibitemOpen
  \bibfield  {author} {\bibinfo {author} {\bibfnamefont {A.~P.}\ \bibnamefont
  {Schnyder}}, \bibinfo {author} {\bibfnamefont {S.}~\bibnamefont {Ryu}},
  \bibinfo {author} {\bibfnamefont {A.}~\bibnamefont {Furusaki}}, \ and\
  \bibinfo {author} {\bibfnamefont {A.~W.~W.}\ \bibnamefont {Ludwig}},\ }\href
  {\doibase 10.1103/PhysRevB.78.195125} {\bibfield  {journal} {\bibinfo
  {journal} {Phys. Rev. B}\ }\textbf {\bibinfo {volume} {78}},\ \bibinfo
  {pages} {195125} (\bibinfo {year} {2008})}\BibitemShut {NoStop}%
\bibitem [{\citenamefont {Ryu}\ \emph {et~al.}(2010)\citenamefont {Ryu},
  \citenamefont {Schnyder}, \citenamefont {Furusaki},\ and\ \citenamefont
  {Ludwig}}]{Ryu10}%
  \BibitemOpen
  \bibfield  {author} {\bibinfo {author} {\bibfnamefont {S.}~\bibnamefont
  {Ryu}}, \bibinfo {author} {\bibfnamefont {A.~P.}\ \bibnamefont {Schnyder}},
  \bibinfo {author} {\bibfnamefont {A.}~\bibnamefont {Furusaki}}, \ and\
  \bibinfo {author} {\bibfnamefont {A.~W.~W.}\ \bibnamefont {Ludwig}},\ }\href
  {http://stacks.iop.org/1367-2630/12/i=6/a=065010} {\bibfield  {journal}
  {\bibinfo  {journal} {New J. Phys.}\ }\textbf {\bibinfo {volume} {12}},\
  \bibinfo {pages} {065010} (\bibinfo {year} {2010})}\BibitemShut {NoStop}%
\bibitem [{\citenamefont {Chiu}\ \emph {et~al.}(2016)\citenamefont {Chiu},
  \citenamefont {Teo}, \citenamefont {Schnyder},\ and\ \citenamefont
  {Ryu}}]{Chiu16}%
  \BibitemOpen
  \bibfield  {author} {\bibinfo {author} {\bibfnamefont {C.-K.}\ \bibnamefont
  {Chiu}}, \bibinfo {author} {\bibfnamefont {J.~C.~Y.}\ \bibnamefont {Teo}},
  \bibinfo {author} {\bibfnamefont {A.~P.}\ \bibnamefont {Schnyder}}, \ and\
  \bibinfo {author} {\bibfnamefont {S.}~\bibnamefont {Ryu}},\ }\href {\doibase
  10.1103/RevModPhys.88.035005} {\bibfield  {journal} {\bibinfo  {journal}
  {Rev. Mod. Phys.}\ }\textbf {\bibinfo {volume} {88}},\ \bibinfo {pages}
  {035005} (\bibinfo {year} {2016})}\BibitemShut {NoStop}%
\bibitem [{\citenamefont {B{\"u}ttiker}(2009)}]{Buttiker09}%
  \BibitemOpen
  \bibfield  {author} {\bibinfo {author} {\bibfnamefont {M.}~\bibnamefont
  {B{\"u}ttiker}},\ }\href {\doibase 10.1126/science.1177157} {\bibfield
  {journal} {\bibinfo  {journal} {Science}\ }\textbf {\bibinfo {volume}
  {325}},\ \bibinfo {pages} {278} (\bibinfo {year} {2009})}\BibitemShut
  {NoStop}%
\bibitem [{\citenamefont {Sonin}(2011)}]{Sonin11}%
  \BibitemOpen
  \bibfield  {author} {\bibinfo {author} {\bibfnamefont {E.~B.}\ \bibnamefont
  {Sonin}},\ }in\ \href {\doibase 10.1117/12.892272} {\emph {\bibinfo
  {booktitle} {Spintronics IV}}},\ Vol.\ \bibinfo {volume} {8100},\ \bibinfo
  {editor} {edited by\ \bibinfo {editor} {\bibfnamefont {H.-J.~M.}\
  \bibnamefont {Drouhin}}, \bibinfo {editor} {\bibfnamefont {J.-E.}\
  \bibnamefont {Wegrowe}}, \ and\ \bibinfo {editor} {\bibfnamefont
  {M.}~\bibnamefont {Razeghi}}},\ \bibinfo {organization} {International
  Society for Optics and Photonics}\ (\bibinfo  {publisher} {SPIE},\ \bibinfo
  {year} {2011})\ pp.\ \bibinfo {pages} {11 -- 21}\BibitemShut {NoStop}%
\bibitem [{\citenamefont {Ando}(2013)}]{Ando13}%
  \BibitemOpen
  \bibfield  {author} {\bibinfo {author} {\bibfnamefont {Y.}~\bibnamefont
  {Ando}},\ }\href {\doibase 10.7566/JPSJ.82.102001} {\bibfield  {journal}
  {\bibinfo  {journal} {‎J. Phys. Soc. Jpn.}\ }\textbf {\bibinfo {volume}
  {82}},\ \bibinfo {pages} {102001} (\bibinfo {year} {2013})}\BibitemShut
  {NoStop}%
\bibitem [{\citenamefont {Maekawa}\ \emph {et~al.}(2017)\citenamefont
  {Maekawa}, \citenamefont {Valenzuela}, \citenamefont {Saitoh},\ and\
  \citenamefont {Kimura}}]{Maekawa17}%
  \BibitemOpen
  \bibfield  {author} {\bibinfo {author} {\bibfnamefont {S.}~\bibnamefont
  {Maekawa}}, \bibinfo {author} {\bibfnamefont {S.~O.}\ \bibnamefont
  {Valenzuela}}, \bibinfo {author} {\bibfnamefont {E.}~\bibnamefont {Saitoh}},
  \ and\ \bibinfo {author} {\bibfnamefont {T.}~\bibnamefont {Kimura}},\
  }\href@noop {} {\emph {\bibinfo {title} {Spin Current}}}\ (\bibinfo
  {publisher} {Oxford University Press},\ \bibinfo {year} {2017})\BibitemShut
  {NoStop}%
\bibitem [{\citenamefont {K{\"o}nig}\ \emph {et~al.}(2007)\citenamefont
  {K{\"o}nig}, \citenamefont {Wiedmann}, \citenamefont {Br{\"u}ne},
  \citenamefont {Roth}, \citenamefont {Buhmann}, \citenamefont {Molenkamp},
  \citenamefont {Qi},\ and\ \citenamefont {Zhang}}]{Konig07}%
  \BibitemOpen
  \bibfield  {author} {\bibinfo {author} {\bibfnamefont {M.}~\bibnamefont
  {K{\"o}nig}}, \bibinfo {author} {\bibfnamefont {S.}~\bibnamefont {Wiedmann}},
  \bibinfo {author} {\bibfnamefont {C.}~\bibnamefont {Br{\"u}ne}}, \bibinfo
  {author} {\bibfnamefont {A.}~\bibnamefont {Roth}}, \bibinfo {author}
  {\bibfnamefont {H.}~\bibnamefont {Buhmann}}, \bibinfo {author} {\bibfnamefont
  {L.~W.}\ \bibnamefont {Molenkamp}}, \bibinfo {author} {\bibfnamefont {X.-L.}\
  \bibnamefont {Qi}}, \ and\ \bibinfo {author} {\bibfnamefont {S.-C.}\
  \bibnamefont {Zhang}},\ }\href {\doibase 10.1126/science.1148047} {\bibfield
  {journal} {\bibinfo  {journal} {Science}\ }\textbf {\bibinfo {volume}
  {318}},\ \bibinfo {pages} {766} (\bibinfo {year} {2007})}\BibitemShut
  {NoStop}%
\bibitem [{\citenamefont {K\"{o}nig}\ \emph {et~al.}(2008)\citenamefont
  {K\"{o}nig}, \citenamefont {Buhmann}, \citenamefont {Molenkamp},
  \citenamefont {Hughes}, \citenamefont {Liu}, \citenamefont {Qi},\ and\
  \citenamefont {Zhang}}]{Konig08}%
  \BibitemOpen
  \bibfield  {author} {\bibinfo {author} {\bibfnamefont {M.}~\bibnamefont
  {K\"{o}nig}}, \bibinfo {author} {\bibfnamefont {H.}~\bibnamefont {Buhmann}},
  \bibinfo {author} {\bibfnamefont {L.~W.}\ \bibnamefont {Molenkamp}}, \bibinfo
  {author} {\bibfnamefont {T.}~\bibnamefont {Hughes}}, \bibinfo {author}
  {\bibfnamefont {C.-X.}\ \bibnamefont {Liu}}, \bibinfo {author} {\bibfnamefont
  {X.-L.}\ \bibnamefont {Qi}}, \ and\ \bibinfo {author} {\bibfnamefont {S.-C.}\
  \bibnamefont {Zhang}},\ }\href {\doibase 10.1143/JPSJ.77.031007} {\bibfield
  {journal} {\bibinfo  {journal} {J. Phys. Soc. Jpn.}\ }\textbf {\bibinfo
  {volume} {77}},\ \bibinfo {pages} {031007} (\bibinfo {year}
  {2008})}\BibitemShut {NoStop}%
\bibitem [{\citenamefont {Hasan}\ and\ \citenamefont {Moore}(2011)}]{Hasan11}%
  \BibitemOpen
  \bibfield  {author} {\bibinfo {author} {\bibfnamefont {M.~Z.}\ \bibnamefont
  {Hasan}}\ and\ \bibinfo {author} {\bibfnamefont {J.~E.}\ \bibnamefont
  {Moore}},\ }\href {\doibase 10.1146/annurev-conmatphys-062910-140432}
  {\bibfield  {journal} {\bibinfo  {journal} {Annu. Rev. Condens. Matter
  Phys.}\ }\textbf {\bibinfo {volume} {2}},\ \bibinfo {pages} {55} (\bibinfo
  {year} {2011})}\BibitemShut {NoStop}%
\bibitem [{\citenamefont {He}\ \emph {et~al.}(2013)\citenamefont {He},
  \citenamefont {Kou},\ and\ \citenamefont {Wang}}]{He13}%
  \BibitemOpen
  \bibfield  {author} {\bibinfo {author} {\bibfnamefont {L.}~\bibnamefont
  {He}}, \bibinfo {author} {\bibfnamefont {X.}~\bibnamefont {Kou}}, \ and\
  \bibinfo {author} {\bibfnamefont {K.~L.}\ \bibnamefont {Wang}},\ }\href
  {\doibase 10.1002/pssr.201307003} {\bibfield  {journal} {\bibinfo  {journal}
  {Phys. Status Solidi RRL}\ }\textbf {\bibinfo {volume} {7}},\ \bibinfo
  {pages} {50} (\bibinfo {year} {2013})}\BibitemShut {NoStop}%
\bibitem [{\citenamefont {Tian}\ \emph {et~al.}(2017)\citenamefont {Tian},
  \citenamefont {Yu}, \citenamefont {Shi},\ and\ \citenamefont
  {Wang}}]{Tian17}%
  \BibitemOpen
  \bibfield  {author} {\bibinfo {author} {\bibfnamefont {W.}~\bibnamefont
  {Tian}}, \bibinfo {author} {\bibfnamefont {W.}~\bibnamefont {Yu}}, \bibinfo
  {author} {\bibfnamefont {J.}~\bibnamefont {Shi}}, \ and\ \bibinfo {author}
  {\bibfnamefont {Y.}~\bibnamefont {Wang}},\ }\href {\doibase
  10.3390/ma10070814} {\bibfield  {journal} {\bibinfo  {journal} {Materials}\
  }\textbf {\bibinfo {volume} {10}},\ \bibinfo {pages} {814} (\bibinfo {year}
  {2017})}\BibitemShut {NoStop}%
\bibitem [{\citenamefont {Hsieh}\ \emph {et~al.}(2009)\citenamefont {Hsieh},
  \citenamefont {Xia}, \citenamefont {Qian}, \citenamefont {Wray},
  \citenamefont {Dil}, \citenamefont {Meier}, \citenamefont {Osterwalder},
  \citenamefont {Patthey}, \citenamefont {Checkelsky}, \citenamefont {Ong},
  \citenamefont {Fedorov}, \citenamefont {Lin}, \citenamefont {Bansil},
  \citenamefont {Grauer}, \citenamefont {Hor}, \citenamefont {Cava},\ and\
  \citenamefont {Hasan}}]{Hsieh09}%
  \BibitemOpen
  \bibfield  {author} {\bibinfo {author} {\bibfnamefont {D.}~\bibnamefont
  {Hsieh}}, \bibinfo {author} {\bibfnamefont {Y.}~\bibnamefont {Xia}}, \bibinfo
  {author} {\bibfnamefont {D.}~\bibnamefont {Qian}}, \bibinfo {author}
  {\bibfnamefont {L.}~\bibnamefont {Wray}}, \bibinfo {author} {\bibfnamefont
  {J.~H.}\ \bibnamefont {Dil}}, \bibinfo {author} {\bibfnamefont
  {F.}~\bibnamefont {Meier}}, \bibinfo {author} {\bibfnamefont
  {J.}~\bibnamefont {Osterwalder}}, \bibinfo {author} {\bibfnamefont
  {L.}~\bibnamefont {Patthey}}, \bibinfo {author} {\bibfnamefont {J.~G.}\
  \bibnamefont {Checkelsky}}, \bibinfo {author} {\bibfnamefont {N.~P.}\
  \bibnamefont {Ong}}, \bibinfo {author} {\bibfnamefont {A.~V.}\ \bibnamefont
  {Fedorov}}, \bibinfo {author} {\bibfnamefont {H.}~\bibnamefont {Lin}},
  \bibinfo {author} {\bibfnamefont {A.}~\bibnamefont {Bansil}}, \bibinfo
  {author} {\bibfnamefont {D.}~\bibnamefont {Grauer}}, \bibinfo {author}
  {\bibfnamefont {Y.~S.}\ \bibnamefont {Hor}}, \bibinfo {author} {\bibfnamefont
  {R.~J.}\ \bibnamefont {Cava}}, \ and\ \bibinfo {author} {\bibfnamefont
  {M.~Z.}\ \bibnamefont {Hasan}},\ }\href {\doibase 10.1038/nature08234}
  {\bibfield  {journal} {\bibinfo  {journal} {Nature}\ }\textbf {\bibinfo
  {volume} {460}},\ \bibinfo {pages} {1101} (\bibinfo {year}
  {2009})}\BibitemShut {NoStop}%
\bibitem [{\citenamefont {Zhang}\ \emph {et~al.}(2011)\citenamefont {Zhang},
  \citenamefont {Chang}, \citenamefont {Zhang}, \citenamefont {Wen},
  \citenamefont {Feng}, \citenamefont {Li}, \citenamefont {Liu}, \citenamefont
  {He}, \citenamefont {Wang}, \citenamefont {Chen}, \citenamefont {Xue},
  \citenamefont {Ma},\ and\ \citenamefont {Wang}}]{Zhang11}%
  \BibitemOpen
  \bibfield  {author} {\bibinfo {author} {\bibfnamefont {J.}~\bibnamefont
  {Zhang}}, \bibinfo {author} {\bibfnamefont {C.-Z.}\ \bibnamefont {Chang}},
  \bibinfo {author} {\bibfnamefont {Z.}~\bibnamefont {Zhang}}, \bibinfo
  {author} {\bibfnamefont {J.}~\bibnamefont {Wen}}, \bibinfo {author}
  {\bibfnamefont {X.}~\bibnamefont {Feng}}, \bibinfo {author} {\bibfnamefont
  {K.}~\bibnamefont {Li}}, \bibinfo {author} {\bibfnamefont {M.}~\bibnamefont
  {Liu}}, \bibinfo {author} {\bibfnamefont {K.}~\bibnamefont {He}}, \bibinfo
  {author} {\bibfnamefont {L.}~\bibnamefont {Wang}}, \bibinfo {author}
  {\bibfnamefont {X.}~\bibnamefont {Chen}}, \bibinfo {author} {\bibfnamefont
  {Q.-K.}\ \bibnamefont {Xue}}, \bibinfo {author} {\bibfnamefont
  {X.}~\bibnamefont {Ma}}, \ and\ \bibinfo {author} {\bibfnamefont
  {Y.}~\bibnamefont {Wang}},\ }\href {\doibase 10.1038/ncomms1588} {\bibfield
  {journal} {\bibinfo  {journal} {Nat. Commun.}\ }\textbf {\bibinfo {volume}
  {2}},\ \bibinfo {pages} {574} (\bibinfo {year} {2011})}\BibitemShut {NoStop}%
\bibitem [{\citenamefont {Kondou}\ \emph {et~al.}(2016)\citenamefont {Kondou},
  \citenamefont {Yoshimi}, \citenamefont {Tsukazaki}, \citenamefont {Fukuma},
  \citenamefont {Matsuno}, \citenamefont {Takahashi}, \citenamefont {Kawasaki},
  \citenamefont {Tokura},\ and\ \citenamefont {Otani}}]{Kondou16}%
  \BibitemOpen
  \bibfield  {author} {\bibinfo {author} {\bibfnamefont {K.}~\bibnamefont
  {Kondou}}, \bibinfo {author} {\bibfnamefont {R.}~\bibnamefont {Yoshimi}},
  \bibinfo {author} {\bibfnamefont {A.}~\bibnamefont {Tsukazaki}}, \bibinfo
  {author} {\bibfnamefont {Y.}~\bibnamefont {Fukuma}}, \bibinfo {author}
  {\bibfnamefont {J.}~\bibnamefont {Matsuno}}, \bibinfo {author} {\bibfnamefont
  {K.~S.}\ \bibnamefont {Takahashi}}, \bibinfo {author} {\bibfnamefont
  {M.}~\bibnamefont {Kawasaki}}, \bibinfo {author} {\bibfnamefont
  {Y.}~\bibnamefont {Tokura}}, \ and\ \bibinfo {author} {\bibfnamefont
  {Y.}~\bibnamefont {Otani}},\ }\href {\doibase 10.1038/nphys3833} {\bibfield
  {journal} {\bibinfo  {journal} {Nat. Phys.}\ }\textbf {\bibinfo {volume}
  {12}},\ \bibinfo {pages} {1027} (\bibinfo {year} {2016})}\BibitemShut
  {NoStop}%
\bibitem [{\citenamefont {Bahramy}\ \emph {et~al.}(2012)\citenamefont
  {Bahramy}, \citenamefont {King}, \citenamefont {de~la Torre}, \citenamefont
  {Chang}, \citenamefont {Shi}, \citenamefont {Patthey}, \citenamefont
  {Balakrishnan}, \citenamefont {Hofmann}, \citenamefont {Arita}, \citenamefont
  {Nagaosa},\ and\ \citenamefont {Baumberger}}]{Bahramy12}%
  \BibitemOpen
  \bibfield  {author} {\bibinfo {author} {\bibfnamefont {M.}~\bibnamefont
  {Bahramy}}, \bibinfo {author} {\bibfnamefont {P.}~\bibnamefont {King}},
  \bibinfo {author} {\bibfnamefont {A.}~\bibnamefont {de~la Torre}}, \bibinfo
  {author} {\bibfnamefont {J.}~\bibnamefont {Chang}}, \bibinfo {author}
  {\bibfnamefont {M.}~\bibnamefont {Shi}}, \bibinfo {author} {\bibfnamefont
  {L.}~\bibnamefont {Patthey}}, \bibinfo {author} {\bibfnamefont
  {G.}~\bibnamefont {Balakrishnan}}, \bibinfo {author} {\bibfnamefont
  {P.}~\bibnamefont {Hofmann}}, \bibinfo {author} {\bibfnamefont
  {R.}~\bibnamefont {Arita}}, \bibinfo {author} {\bibfnamefont
  {N.}~\bibnamefont {Nagaosa}}, \ and\ \bibinfo {author} {\bibfnamefont
  {F.}~\bibnamefont {Baumberger}},\ }\href {\doibase 10.1038/ncomms2162}
  {\bibfield  {journal} {\bibinfo  {journal} {Nat. Phys.}\ }\textbf {\bibinfo
  {volume} {3}},\ \bibinfo {pages} {1159} (\bibinfo {year} {2012})}\BibitemShut
  {NoStop}%
\bibitem [{\citenamefont {Beidenkopf}\ \emph {et~al.}(2011)\citenamefont
  {Beidenkopf}, \citenamefont {Roushan}, \citenamefont {Seo}, \citenamefont
  {Gorman}, \citenamefont {Drozdov}, \citenamefont {Hor}, \citenamefont
  {Cava},\ and\ \citenamefont {Yazdani}}]{Beidenkopf11}%
  \BibitemOpen
  \bibfield  {author} {\bibinfo {author} {\bibfnamefont {H.}~\bibnamefont
  {Beidenkopf}}, \bibinfo {author} {\bibfnamefont {P.}~\bibnamefont {Roushan}},
  \bibinfo {author} {\bibfnamefont {J.}~\bibnamefont {Seo}}, \bibinfo {author}
  {\bibfnamefont {L.}~\bibnamefont {Gorman}}, \bibinfo {author} {\bibfnamefont
  {I.}~\bibnamefont {Drozdov}}, \bibinfo {author} {\bibfnamefont {Y.~S.}\
  \bibnamefont {Hor}}, \bibinfo {author} {\bibfnamefont {R.~J.}\ \bibnamefont
  {Cava}}, \ and\ \bibinfo {author} {\bibfnamefont {A.}~\bibnamefont
  {Yazdani}},\ }\href {\doibase 10.1038/nphys2108} {\bibfield  {journal}
  {\bibinfo  {journal} {Nat. Phys.}\ }\textbf {\bibinfo {volume} {7}},\
  \bibinfo {pages} {939} (\bibinfo {year} {2011})}\BibitemShut {NoStop}%
\bibitem [{\citenamefont {Akhmerov}\ and\ \citenamefont
  {Beenakker}(2008)}]{Akhmerov08}%
  \BibitemOpen
  \bibfield  {author} {\bibinfo {author} {\bibfnamefont {A.~R.}\ \bibnamefont
  {Akhmerov}}\ and\ \bibinfo {author} {\bibfnamefont {C.~W.~J.}\ \bibnamefont
  {Beenakker}},\ }\href {\doibase 10.1103/PhysRevB.77.085423} {\bibfield
  {journal} {\bibinfo  {journal} {Phys. Rev. B}\ }\textbf {\bibinfo {volume}
  {77}},\ \bibinfo {pages} {085423} (\bibinfo {year} {2008})}\BibitemShut
  {NoStop}%
\bibitem [{\citenamefont {Cho}\ \emph {et~al.}(2012)\citenamefont {Cho},
  \citenamefont {Kim}, \citenamefont {Syers}, \citenamefont {Butch},
  \citenamefont {Paglione},\ and\ \citenamefont {Fuhrer}}]{Cho12}%
  \BibitemOpen
  \bibfield  {author} {\bibinfo {author} {\bibfnamefont {S.}~\bibnamefont
  {Cho}}, \bibinfo {author} {\bibfnamefont {D.}~\bibnamefont {Kim}}, \bibinfo
  {author} {\bibfnamefont {P.}~\bibnamefont {Syers}}, \bibinfo {author}
  {\bibfnamefont {N.~P.}\ \bibnamefont {Butch}}, \bibinfo {author}
  {\bibfnamefont {J.}~\bibnamefont {Paglione}}, \ and\ \bibinfo {author}
  {\bibfnamefont {M.~S.}\ \bibnamefont {Fuhrer}},\ }\href {\doibase
  10.1021/nl203851g} {\bibfield  {journal} {\bibinfo  {journal} {Nano Lett.}\
  }\textbf {\bibinfo {volume} {12}},\ \bibinfo {pages} {469} (\bibinfo {year}
  {2012})}\BibitemShut {NoStop}%
\bibitem [{\citenamefont {Claro}\ \emph {et~al.}(2019)\citenamefont {Claro},
  \citenamefont {Levy}, \citenamefont {Gangopadhyay}, \citenamefont {Smith},\
  and\ \citenamefont {Tamargo}}]{Claro19}%
  \BibitemOpen
  \bibfield  {author} {\bibinfo {author} {\bibfnamefont {M.~S.}\ \bibnamefont
  {Claro}}, \bibinfo {author} {\bibfnamefont {I.}~\bibnamefont {Levy}},
  \bibinfo {author} {\bibfnamefont {A.}~\bibnamefont {Gangopadhyay}}, \bibinfo
  {author} {\bibfnamefont {D.~J.}\ \bibnamefont {Smith}}, \ and\ \bibinfo
  {author} {\bibfnamefont {M.~C.}\ \bibnamefont {Tamargo}},\ }\href {\doibase
  10.1038/s41598-019-39821-y} {\bibfield  {journal} {\bibinfo  {journal} {Sci.
  Rep.}\ }\textbf {\bibinfo {volume} {9}},\ \bibinfo {pages} {3370} (\bibinfo
  {year} {2019})}\BibitemShut {NoStop}%
\bibitem [{\citenamefont {Jing}\ \emph {et~al.}(2019)\citenamefont {Jing},
  \citenamefont {Huang}, \citenamefont {Wu}, \citenamefont {Meng},
  \citenamefont {Li}, \citenamefont {Zhou}, \citenamefont {Peng},\ and\
  \citenamefont {Xu}}]{Jing19}%
  \BibitemOpen
  \bibfield  {author} {\bibinfo {author} {\bibfnamefont {Y.}~\bibnamefont
  {Jing}}, \bibinfo {author} {\bibfnamefont {S.}~\bibnamefont {Huang}},
  \bibinfo {author} {\bibfnamefont {J.}~\bibnamefont {Wu}}, \bibinfo {author}
  {\bibfnamefont {M.}~\bibnamefont {Meng}}, \bibinfo {author} {\bibfnamefont
  {X.}~\bibnamefont {Li}}, \bibinfo {author} {\bibfnamefont {Y.}~\bibnamefont
  {Zhou}}, \bibinfo {author} {\bibfnamefont {H.}~\bibnamefont {Peng}}, \ and\
  \bibinfo {author} {\bibfnamefont {H.}~\bibnamefont {Xu}},\ }\href {\doibase
  10.1002/adma.201903686} {\bibfield  {journal} {\bibinfo  {journal} {Adv.
  Mater.}\ }\textbf {\bibinfo {volume} {31}},\ \bibinfo {pages} {1903686}
  (\bibinfo {year} {2019})}\BibitemShut {NoStop}%
\bibitem [{\citenamefont {Huang}\ \emph {et~al.}(2019)\citenamefont {Huang},
  \citenamefont {Lou}, \citenamefont {Cheng}, \citenamefont {Yang},\ and\
  \citenamefont {Chang}}]{Huang19}%
  \BibitemOpen
  \bibfield  {author} {\bibinfo {author} {\bibfnamefont {Y.}~\bibnamefont
  {Huang}}, \bibinfo {author} {\bibfnamefont {W.}~\bibnamefont {Lou}}, \bibinfo
  {author} {\bibfnamefont {F.}~\bibnamefont {Cheng}}, \bibinfo {author}
  {\bibfnamefont {W.}~\bibnamefont {Yang}}, \ and\ \bibinfo {author}
  {\bibfnamefont {K.}~\bibnamefont {Chang}},\ }\href {\doibase
  10.1103/PhysRevApplied.12.034003} {\bibfield  {journal} {\bibinfo  {journal}
  {Phys. Rev. Applied}\ }\textbf {\bibinfo {volume} {12}},\ \bibinfo {pages}
  {034003} (\bibinfo {year} {2019})}\BibitemShut {NoStop}%
\bibitem [{\citenamefont {Chen}\ and\ \citenamefont
  {Schnyder}(2019)}]{Chen19_universality_class}%
  \BibitemOpen
  \bibfield  {author} {\bibinfo {author} {\bibfnamefont {W.}~\bibnamefont
  {Chen}}\ and\ \bibinfo {author} {\bibfnamefont {A.~P.}\ \bibnamefont
  {Schnyder}},\ }\href {\doibase 10.1088/1367-2630/ab2a2d} {\bibfield
  {journal} {\bibinfo  {journal} {New J. Phys.}\ }\textbf {\bibinfo {volume}
  {21}},\ \bibinfo {pages} {073003} (\bibinfo {year} {2019})}\BibitemShut
  {NoStop}%
\bibitem [{\citenamefont {Ara\'ujo}\ \emph {et~al.}(2019)\citenamefont
  {Ara\'ujo}, \citenamefont {Maciel}, \citenamefont {Dornelas}, \citenamefont
  {Varjas},\ and\ \citenamefont {Ferreira}}]{Araujo19}%
  \BibitemOpen
  \bibfield  {author} {\bibinfo {author} {\bibfnamefont {A.~L.}\ \bibnamefont
  {Ara\'ujo}}, \bibinfo {author} {\bibfnamefont {R.~P.}\ \bibnamefont
  {Maciel}}, \bibinfo {author} {\bibfnamefont {R.~G.~F.}\ \bibnamefont
  {Dornelas}}, \bibinfo {author} {\bibfnamefont {D.}~\bibnamefont {Varjas}}, \
  and\ \bibinfo {author} {\bibfnamefont {G.~J.}\ \bibnamefont {Ferreira}},\
  }\href {\doibase 10.1103/PhysRevB.100.205111} {\bibfield  {journal} {\bibinfo
   {journal} {Phys. Rev. B}\ }\textbf {\bibinfo {volume} {100}},\ \bibinfo
  {pages} {205111} (\bibinfo {year} {2019})}\BibitemShut {NoStop}%
\bibitem [{\citenamefont {Halperin}(1982)}]{Halperin82}%
  \BibitemOpen
  \bibfield  {author} {\bibinfo {author} {\bibfnamefont {B.~I.}\ \bibnamefont
  {Halperin}},\ }\href {\doibase 10.1103/PhysRevB.25.2185} {\bibfield
  {journal} {\bibinfo  {journal} {Phys. Rev. B}\ }\textbf {\bibinfo {volume}
  {25}},\ \bibinfo {pages} {2185} (\bibinfo {year} {1982})}\BibitemShut
  {NoStop}%
\bibitem [{\citenamefont {Smrcka}(1984)}]{Smrcka84}%
  \BibitemOpen
  \bibfield  {author} {\bibinfo {author} {\bibfnamefont {L.}~\bibnamefont
  {Smrcka}},\ }\href {\doibase 10.1088/0022-3719/17/2/006} {\bibfield
  {journal} {\bibinfo  {journal} {J. Phys. Condens. Matter}\ }\textbf {\bibinfo
  {volume} {17}},\ \bibinfo {pages} {L63} (\bibinfo {year} {1984})}\BibitemShut
  {NoStop}%
\bibitem [{\citenamefont {MacDonald}\ and\ \citenamefont
  {St\ifmmode~\check{r}\else \v{r}\fi{}eda}(1984)}]{MacDonald84}%
  \BibitemOpen
  \bibfield  {author} {\bibinfo {author} {\bibfnamefont {A.~H.}\ \bibnamefont
  {MacDonald}}\ and\ \bibinfo {author} {\bibfnamefont {P.}~\bibnamefont
  {St\ifmmode~\check{r}\else \v{r}\fi{}eda}},\ }\href {\doibase
  10.1103/PhysRevB.29.1616} {\bibfield  {journal} {\bibinfo  {journal} {Phys.
  Rev. B}\ }\textbf {\bibinfo {volume} {29}},\ \bibinfo {pages} {1616}
  (\bibinfo {year} {1984})}\BibitemShut {NoStop}%
\bibitem [{\citenamefont {Streda}\ \emph {et~al.}(1987)\citenamefont {Streda},
  \citenamefont {Kucera},\ and\ \citenamefont {MacDonald}}]{Streda87}%
  \BibitemOpen
  \bibfield  {author} {\bibinfo {author} {\bibfnamefont {P.}~\bibnamefont
  {Streda}}, \bibinfo {author} {\bibfnamefont {J.}~\bibnamefont {Kucera}}, \
  and\ \bibinfo {author} {\bibfnamefont {A.~H.}\ \bibnamefont {MacDonald}},\
  }\href {\doibase 10.1103/PhysRevLett.59.1973} {\bibfield  {journal} {\bibinfo
   {journal} {Phys. Rev. Lett.}\ }\textbf {\bibinfo {volume} {59}},\ \bibinfo
  {pages} {1973} (\bibinfo {year} {1987})}\BibitemShut {NoStop}%
\bibitem [{\citenamefont {B\"uttiker}(1988)}]{Buttiker88}%
  \BibitemOpen
  \bibfield  {author} {\bibinfo {author} {\bibfnamefont {M.}~\bibnamefont
  {B\"uttiker}},\ }\href {\doibase 10.1103/PhysRevB.38.9375} {\bibfield
  {journal} {\bibinfo  {journal} {Phys. Rev. B}\ }\textbf {\bibinfo {volume}
  {38}},\ \bibinfo {pages} {9375} (\bibinfo {year} {1988})}\BibitemShut
  {NoStop}%
\bibitem [{\citenamefont {Chklovskii}\ \emph {et~al.}(1992)\citenamefont
  {Chklovskii}, \citenamefont {Shklovskii},\ and\ \citenamefont
  {Glazman}}]{Chklovskii92}%
  \BibitemOpen
  \bibfield  {author} {\bibinfo {author} {\bibfnamefont {D.~B.}\ \bibnamefont
  {Chklovskii}}, \bibinfo {author} {\bibfnamefont {B.~I.}\ \bibnamefont
  {Shklovskii}}, \ and\ \bibinfo {author} {\bibfnamefont {L.~I.}\ \bibnamefont
  {Glazman}},\ }\href {\doibase 10.1103/PhysRevB.46.4026} {\bibfield  {journal}
  {\bibinfo  {journal} {Phys. Rev. B}\ }\textbf {\bibinfo {volume} {46}},\
  \bibinfo {pages} {4026} (\bibinfo {year} {1992})}\BibitemShut {NoStop}%
\bibitem [{\citenamefont {Chang}\ \emph {et~al.}(2013)\citenamefont {Chang},
  \citenamefont {Zhang}, \citenamefont {Feng}, \citenamefont {Shen},
  \citenamefont {Zhang}, \citenamefont {Guo}, \citenamefont {Li}, \citenamefont
  {Ou}, \citenamefont {Wei}, \citenamefont {Wang}, \citenamefont {Ji},
  \citenamefont {Feng}, \citenamefont {Ji}, \citenamefont {Chen}, \citenamefont
  {Jia}, \citenamefont {Dai}, \citenamefont {Fang}, \citenamefont {Zhang},
  \citenamefont {He}, \citenamefont {Wang}, \citenamefont {Lu}, \citenamefont
  {Ma},\ and\ \citenamefont {Xue}}]{Chang13}%
  \BibitemOpen
  \bibfield  {author} {\bibinfo {author} {\bibfnamefont {C.-Z.}\ \bibnamefont
  {Chang}}, \bibinfo {author} {\bibfnamefont {J.}~\bibnamefont {Zhang}},
  \bibinfo {author} {\bibfnamefont {X.}~\bibnamefont {Feng}}, \bibinfo {author}
  {\bibfnamefont {J.}~\bibnamefont {Shen}}, \bibinfo {author} {\bibfnamefont
  {Z.}~\bibnamefont {Zhang}}, \bibinfo {author} {\bibfnamefont
  {M.}~\bibnamefont {Guo}}, \bibinfo {author} {\bibfnamefont {K.}~\bibnamefont
  {Li}}, \bibinfo {author} {\bibfnamefont {Y.}~\bibnamefont {Ou}}, \bibinfo
  {author} {\bibfnamefont {P.}~\bibnamefont {Wei}}, \bibinfo {author}
  {\bibfnamefont {L.-L.}\ \bibnamefont {Wang}}, \bibinfo {author}
  {\bibfnamefont {Z.-Q.}\ \bibnamefont {Ji}}, \bibinfo {author} {\bibfnamefont
  {Y.}~\bibnamefont {Feng}}, \bibinfo {author} {\bibfnamefont {S.}~\bibnamefont
  {Ji}}, \bibinfo {author} {\bibfnamefont {X.}~\bibnamefont {Chen}}, \bibinfo
  {author} {\bibfnamefont {J.}~\bibnamefont {Jia}}, \bibinfo {author}
  {\bibfnamefont {X.}~\bibnamefont {Dai}}, \bibinfo {author} {\bibfnamefont
  {Z.}~\bibnamefont {Fang}}, \bibinfo {author} {\bibfnamefont {S.-C.}\
  \bibnamefont {Zhang}}, \bibinfo {author} {\bibfnamefont {K.}~\bibnamefont
  {He}}, \bibinfo {author} {\bibfnamefont {Y.}~\bibnamefont {Wang}}, \bibinfo
  {author} {\bibfnamefont {L.}~\bibnamefont {Lu}}, \bibinfo {author}
  {\bibfnamefont {X.-C.}\ \bibnamefont {Ma}}, \ and\ \bibinfo {author}
  {\bibfnamefont {Q.-K.}\ \bibnamefont {Xue}},\ }\href {\doibase
  10.1126/science.1234414} {\bibfield  {journal} {\bibinfo  {journal}
  {Science}\ }\textbf {\bibinfo {volume} {340}},\ \bibinfo {pages} {167}
  (\bibinfo {year} {2013})}\BibitemShut {NoStop}%
\bibitem [{\citenamefont {Liu}\ \emph {et~al.}(2016)\citenamefont {Liu},
  \citenamefont {Zhang},\ and\ \citenamefont {Qi}}]{Liu16}%
  \BibitemOpen
  \bibfield  {author} {\bibinfo {author} {\bibfnamefont {C.-X.}\ \bibnamefont
  {Liu}}, \bibinfo {author} {\bibfnamefont {S.-C.}\ \bibnamefont {Zhang}}, \
  and\ \bibinfo {author} {\bibfnamefont {X.-L.}\ \bibnamefont {Qi}},\ }\href
  {\doibase 10.1146/annurev-conmatphys-031115-011417} {\bibfield  {journal}
  {\bibinfo  {journal} {Annu. Rev. Condens. Matter Phys.}\ }\textbf {\bibinfo
  {volume} {7}},\ \bibinfo {pages} {301} (\bibinfo {year} {2016})}\BibitemShut
  {NoStop}%
\bibitem [{\citenamefont {He}\ \emph {et~al.}(2018)\citenamefont {He},
  \citenamefont {Wang},\ and\ \citenamefont {Xue}}]{He18}%
  \BibitemOpen
  \bibfield  {author} {\bibinfo {author} {\bibfnamefont {K.}~\bibnamefont
  {He}}, \bibinfo {author} {\bibfnamefont {Y.}~\bibnamefont {Wang}}, \ and\
  \bibinfo {author} {\bibfnamefont {Q.-K.}\ \bibnamefont {Xue}},\ }\href
  {\doibase 10.1146/annurev-conmatphys-033117-054144} {\bibfield  {journal}
  {\bibinfo  {journal} {Annu. Rev. Condens. Matter Phys.}\ }\textbf {\bibinfo
  {volume} {9}},\ \bibinfo {pages} {329} (\bibinfo {year} {2018})}\BibitemShut
  {NoStop}%
\bibitem [{\citenamefont {Ferreira}\ and\ \citenamefont
  {Loss}(2013)}]{Ferreira13}%
  \BibitemOpen
  \bibfield  {author} {\bibinfo {author} {\bibfnamefont {G.~J.}\ \bibnamefont
  {Ferreira}}\ and\ \bibinfo {author} {\bibfnamefont {D.}~\bibnamefont
  {Loss}},\ }\href {\doibase 10.1103/PhysRevLett.111.106802} {\bibfield
  {journal} {\bibinfo  {journal} {Phys. Rev. Lett.}\ }\textbf {\bibinfo
  {volume} {111}},\ \bibinfo {pages} {106802} (\bibinfo {year}
  {2013})}\BibitemShut {NoStop}%
\bibitem [{\citenamefont {Bernevig}\ \emph {et~al.}(2006)\citenamefont
  {Bernevig}, \citenamefont {Hughes},\ and\ \citenamefont
  {Zhang}}]{Bernevig06}%
  \BibitemOpen
  \bibfield  {author} {\bibinfo {author} {\bibfnamefont {B.~A.}\ \bibnamefont
  {Bernevig}}, \bibinfo {author} {\bibfnamefont {T.~L.}\ \bibnamefont
  {Hughes}}, \ and\ \bibinfo {author} {\bibfnamefont {S.-C.}\ \bibnamefont
  {Zhang}},\ }\href {\doibase 10.1126/science.1133734} {\bibfield  {journal}
  {\bibinfo  {journal} {Science}\ }\textbf {\bibinfo {volume} {314}},\ \bibinfo
  {pages} {1757} (\bibinfo {year} {2006})}\BibitemShut {NoStop}%
\bibitem [{\citenamefont {Hirsch}(1990)}]{Hirsch90}%
  \BibitemOpen
  \bibfield  {author} {\bibinfo {author} {\bibfnamefont {J.~E.}\ \bibnamefont
  {Hirsch}},\ }\href {\doibase 10.1103/PhysRevB.42.4774} {\bibfield  {journal}
  {\bibinfo  {journal} {Phys. Rev. B}\ }\textbf {\bibinfo {volume} {42}},\
  \bibinfo {pages} {4774} (\bibinfo {year} {1990})}\BibitemShut {NoStop}%
\bibitem [{\citenamefont {Hirsch}(1999)}]{Hirsch99_2}%
  \BibitemOpen
  \bibfield  {author} {\bibinfo {author} {\bibfnamefont {J.~E.}\ \bibnamefont
  {Hirsch}},\ }\href {\doibase 10.1103/PhysRevB.60.14787} {\bibfield  {journal}
  {\bibinfo  {journal} {Phys. Rev. B}\ }\textbf {\bibinfo {volume} {60}},\
  \bibinfo {pages} {14787} (\bibinfo {year} {1999})}\BibitemShut {NoStop}%
\bibitem [{\citenamefont {Meier}\ and\ \citenamefont {Loss}(2003)}]{Meier03}%
  \BibitemOpen
  \bibfield  {author} {\bibinfo {author} {\bibfnamefont {F.}~\bibnamefont
  {Meier}}\ and\ \bibinfo {author} {\bibfnamefont {D.}~\bibnamefont {Loss}},\
  }\href {\doibase 10.1103/PhysRevLett.90.167204} {\bibfield  {journal}
  {\bibinfo  {journal} {Phys. Rev. Lett.}\ }\textbf {\bibinfo {volume} {90}},\
  \bibinfo {pages} {167204} (\bibinfo {year} {2003})}\BibitemShut {NoStop}%
\bibitem [{\citenamefont {Sch\"utz}\ \emph {et~al.}(2003)\citenamefont
  {Sch\"utz}, \citenamefont {Kollar},\ and\ \citenamefont
  {Kopietz}}]{Schutz03}%
  \BibitemOpen
  \bibfield  {author} {\bibinfo {author} {\bibfnamefont {F.}~\bibnamefont
  {Sch\"utz}}, \bibinfo {author} {\bibfnamefont {M.}~\bibnamefont {Kollar}}, \
  and\ \bibinfo {author} {\bibfnamefont {P.}~\bibnamefont {Kopietz}},\ }\href
  {\doibase 10.1103/PhysRevLett.91.017205} {\bibfield  {journal} {\bibinfo
  {journal} {Phys. Rev. Lett.}\ }\textbf {\bibinfo {volume} {91}},\ \bibinfo
  {pages} {017205} (\bibinfo {year} {2003})}\BibitemShut {NoStop}%
\bibitem [{\citenamefont {Sun}\ \emph {et~al.}(2004)\citenamefont {Sun},
  \citenamefont {Guo},\ and\ \citenamefont {Wang}}]{Sun04}%
  \BibitemOpen
  \bibfield  {author} {\bibinfo {author} {\bibfnamefont {Q.-f.}\ \bibnamefont
  {Sun}}, \bibinfo {author} {\bibfnamefont {H.}~\bibnamefont {Guo}}, \ and\
  \bibinfo {author} {\bibfnamefont {J.}~\bibnamefont {Wang}},\ }\href {\doibase
  10.1103/PhysRevB.69.054409} {\bibfield  {journal} {\bibinfo  {journal} {Phys.
  Rev. B}\ }\textbf {\bibinfo {volume} {69}},\ \bibinfo {pages} {054409}
  (\bibinfo {year} {2004})}\BibitemShut {NoStop}%
\bibitem [{\citenamefont {Chen}(2014)}]{Chen14_inductance_spin_current}%
  \BibitemOpen
  \bibfield  {author} {\bibinfo {author} {\bibfnamefont {W.}~\bibnamefont
  {Chen}},\ }\href {\doibase 10.1063/1.4868543} {\bibfield  {journal} {\bibinfo
   {journal} {J. Appl. Phys.}\ }\textbf {\bibinfo {volume} {115}},\ \bibinfo
  {pages} {113901} (\bibinfo {year} {2014})}\BibitemShut {NoStop}%
\bibitem [{\citenamefont {Candido}\ \emph {et~al.}(2018)\citenamefont
  {Candido}, \citenamefont {Flatt\'e},\ and\ \citenamefont
  {Egues}}]{Candido18}%
  \BibitemOpen
  \bibfield  {author} {\bibinfo {author} {\bibfnamefont {D.~R.}\ \bibnamefont
  {Candido}}, \bibinfo {author} {\bibfnamefont {M.~E.}\ \bibnamefont
  {Flatt\'e}}, \ and\ \bibinfo {author} {\bibfnamefont {J.~C.}\ \bibnamefont
  {Egues}},\ }\href {\doibase 10.1103/PhysRevLett.121.256804} {\bibfield
  {journal} {\bibinfo  {journal} {Phys. Rev. Lett.}\ }\textbf {\bibinfo
  {volume} {121}},\ \bibinfo {pages} {256804} (\bibinfo {year}
  {2018})}\BibitemShut {NoStop}%
\bibitem [{\citenamefont {Qian}\ \emph {et~al.}(2014)\citenamefont {Qian},
  \citenamefont {Liu}, \citenamefont {Fu},\ and\ \citenamefont {Li}}]{Qian14}%
  \BibitemOpen
  \bibfield  {author} {\bibinfo {author} {\bibfnamefont {X.}~\bibnamefont
  {Qian}}, \bibinfo {author} {\bibfnamefont {J.}~\bibnamefont {Liu}}, \bibinfo
  {author} {\bibfnamefont {L.}~\bibnamefont {Fu}}, \ and\ \bibinfo {author}
  {\bibfnamefont {J.}~\bibnamefont {Li}},\ }\href {\doibase
  10.1126/science.1256815} {\bibfield  {journal} {\bibinfo  {journal}
  {Science}\ }\textbf {\bibinfo {volume} {346}},\ \bibinfo {pages} {1344}
  (\bibinfo {year} {2014})}\BibitemShut {NoStop}%
\bibitem [{\citenamefont {Zheng}\ \emph {et~al.}(2016)\citenamefont {Zheng},
  \citenamefont {Cai}, \citenamefont {Ge}, \citenamefont {Zhang}, \citenamefont
  {Liu}, \citenamefont {Lu}, \citenamefont {Zhang}, \citenamefont {Qiu},
  \citenamefont {Taniguchi}, \citenamefont {Watanabe}, \citenamefont {Jia},
  \citenamefont {Qi}, \citenamefont {Chen}, \citenamefont {Sun},\ and\
  \citenamefont {Feng}}]{Zheng16}%
  \BibitemOpen
  \bibfield  {author} {\bibinfo {author} {\bibfnamefont {F.}~\bibnamefont
  {Zheng}}, \bibinfo {author} {\bibfnamefont {C.}~\bibnamefont {Cai}}, \bibinfo
  {author} {\bibfnamefont {S.}~\bibnamefont {Ge}}, \bibinfo {author}
  {\bibfnamefont {X.}~\bibnamefont {Zhang}}, \bibinfo {author} {\bibfnamefont
  {X.}~\bibnamefont {Liu}}, \bibinfo {author} {\bibfnamefont {H.}~\bibnamefont
  {Lu}}, \bibinfo {author} {\bibfnamefont {Y.}~\bibnamefont {Zhang}}, \bibinfo
  {author} {\bibfnamefont {J.}~\bibnamefont {Qiu}}, \bibinfo {author}
  {\bibfnamefont {T.}~\bibnamefont {Taniguchi}}, \bibinfo {author}
  {\bibfnamefont {K.}~\bibnamefont {Watanabe}}, \bibinfo {author}
  {\bibfnamefont {S.}~\bibnamefont {Jia}}, \bibinfo {author} {\bibfnamefont
  {J.}~\bibnamefont {Qi}}, \bibinfo {author} {\bibfnamefont {J.-H.}\
  \bibnamefont {Chen}}, \bibinfo {author} {\bibfnamefont {D.}~\bibnamefont
  {Sun}}, \ and\ \bibinfo {author} {\bibfnamefont {J.}~\bibnamefont {Feng}},\
  }\href {\doibase 10.1002/adma.201600100} {\bibfield  {journal} {\bibinfo
  {journal} {Adv. Mater.}\ }\textbf {\bibinfo {volume} {28}},\ \bibinfo {pages}
  {4845} (\bibinfo {year} {2016})}\BibitemShut {NoStop}%
\bibitem [{\citenamefont {Tang}\ \emph {et~al.}(2017)\citenamefont {Tang},
  \citenamefont {Zhang}, \citenamefont {Wong}, \citenamefont {Pedramrazi},
  \citenamefont {Tsai}, \citenamefont {Jia}, \citenamefont {Moritz},
  \citenamefont {Claassen}, \citenamefont {Ryu}, \citenamefont {Kahn},
  \citenamefont {Jiang}, \citenamefont {Yan}, \citenamefont {Hashimoto},
  \citenamefont {Lu}, \citenamefont {Moore}, \citenamefont {Hwang},
  \citenamefont {Hwang}, \citenamefont {Chen}, \citenamefont {Ugeda},
  \citenamefont {Liu}, \citenamefont {Xie}, \citenamefont {Devereaux},
  \citenamefont {Crommie}, \citenamefont {Mo},\ and\ \citenamefont
  {Shen}}]{Tang17}%
  \BibitemOpen
  \bibfield  {author} {\bibinfo {author} {\bibfnamefont {S.}~\bibnamefont
  {Tang}}, \bibinfo {author} {\bibfnamefont {C.}~\bibnamefont {Zhang}},
  \bibinfo {author} {\bibfnamefont {D.}~\bibnamefont {Wong}}, \bibinfo {author}
  {\bibfnamefont {Z.}~\bibnamefont {Pedramrazi}}, \bibinfo {author}
  {\bibfnamefont {H.-Z.}\ \bibnamefont {Tsai}}, \bibinfo {author}
  {\bibfnamefont {C.}~\bibnamefont {Jia}}, \bibinfo {author} {\bibfnamefont
  {B.}~\bibnamefont {Moritz}}, \bibinfo {author} {\bibfnamefont
  {M.}~\bibnamefont {Claassen}}, \bibinfo {author} {\bibfnamefont
  {H.}~\bibnamefont {Ryu}}, \bibinfo {author} {\bibfnamefont {S.}~\bibnamefont
  {Kahn}}, \bibinfo {author} {\bibfnamefont {J.}~\bibnamefont {Jiang}},
  \bibinfo {author} {\bibfnamefont {H.}~\bibnamefont {Yan}}, \bibinfo {author}
  {\bibfnamefont {M.}~\bibnamefont {Hashimoto}}, \bibinfo {author}
  {\bibfnamefont {D.}~\bibnamefont {Lu}}, \bibinfo {author} {\bibfnamefont
  {R.~G.}\ \bibnamefont {Moore}}, \bibinfo {author} {\bibfnamefont {C.-C.}\
  \bibnamefont {Hwang}}, \bibinfo {author} {\bibfnamefont {Z.}~\bibnamefont
  {Hwang}, \bibfnamefont {Choongyuand~Hussain}}, \bibinfo {author}
  {\bibfnamefont {Y.}~\bibnamefont {Chen}}, \bibinfo {author} {\bibfnamefont
  {M.~M.}\ \bibnamefont {Ugeda}}, \bibinfo {author} {\bibfnamefont
  {Z.}~\bibnamefont {Liu}}, \bibinfo {author} {\bibfnamefont {X.}~\bibnamefont
  {Xie}}, \bibinfo {author} {\bibfnamefont {T.~P.}\ \bibnamefont {Devereaux}},
  \bibinfo {author} {\bibfnamefont {M.~F.}\ \bibnamefont {Crommie}}, \bibinfo
  {author} {\bibfnamefont {S.-K.}\ \bibnamefont {Mo}}, \ and\ \bibinfo {author}
  {\bibfnamefont {Z.-X.}\ \bibnamefont {Shen}},\ }\href {\doibase
  10.1038/nphys4174} {\bibfield  {journal} {\bibinfo  {journal} {Nat. Phys.}\
  }\textbf {\bibinfo {volume} {13}},\ \bibinfo {pages} {683} (\bibinfo {year}
  {2017})}\BibitemShut {NoStop}%
\bibitem [{\citenamefont {Chen}\ \emph {et~al.}(2009)\citenamefont {Chen},
  \citenamefont {Analytis}, \citenamefont {Chu}, \citenamefont {Liu},
  \citenamefont {Mo}, \citenamefont {Qi}, \citenamefont {Zhang}, \citenamefont
  {Lu}, \citenamefont {Dai}, \citenamefont {Fang}, \citenamefont {Zhang},
  \citenamefont {Fisher}, \citenamefont {Hussain},\ and\ \citenamefont
  {Shen}}]{Chen09}%
  \BibitemOpen
  \bibfield  {author} {\bibinfo {author} {\bibfnamefont {Y.~L.}\ \bibnamefont
  {Chen}}, \bibinfo {author} {\bibfnamefont {J.~G.}\ \bibnamefont {Analytis}},
  \bibinfo {author} {\bibfnamefont {J.-H.}\ \bibnamefont {Chu}}, \bibinfo
  {author} {\bibfnamefont {Z.~K.}\ \bibnamefont {Liu}}, \bibinfo {author}
  {\bibfnamefont {S.-K.}\ \bibnamefont {Mo}}, \bibinfo {author} {\bibfnamefont
  {X.~L.}\ \bibnamefont {Qi}}, \bibinfo {author} {\bibfnamefont {H.~J.}\
  \bibnamefont {Zhang}}, \bibinfo {author} {\bibfnamefont {D.~H.}\ \bibnamefont
  {Lu}}, \bibinfo {author} {\bibfnamefont {X.}~\bibnamefont {Dai}}, \bibinfo
  {author} {\bibfnamefont {Z.}~\bibnamefont {Fang}}, \bibinfo {author}
  {\bibfnamefont {S.~C.}\ \bibnamefont {Zhang}}, \bibinfo {author}
  {\bibfnamefont {I.~R.}\ \bibnamefont {Fisher}}, \bibinfo {author}
  {\bibfnamefont {Z.}~\bibnamefont {Hussain}}, \ and\ \bibinfo {author}
  {\bibfnamefont {Z.-X.}\ \bibnamefont {Shen}},\ }\href {\doibase
  10.1126/science.1173034} {\bibfield  {journal} {\bibinfo  {journal}
  {Science}\ }\textbf {\bibinfo {volume} {325}},\ \bibinfo {pages} {178}
  (\bibinfo {year} {2009})}\BibitemShut {NoStop}%
\bibitem [{\citenamefont {Analytis}\ \emph {et~al.}(2010)\citenamefont
  {Analytis}, \citenamefont {Chu}, \citenamefont {Chen}, \citenamefont
  {Corredor}, \citenamefont {McDonald}, \citenamefont {Shen},\ and\
  \citenamefont {Fisher}}]{Analytis10}%
  \BibitemOpen
  \bibfield  {author} {\bibinfo {author} {\bibfnamefont {J.~G.}\ \bibnamefont
  {Analytis}}, \bibinfo {author} {\bibfnamefont {J.-H.}\ \bibnamefont {Chu}},
  \bibinfo {author} {\bibfnamefont {Y.}~\bibnamefont {Chen}}, \bibinfo {author}
  {\bibfnamefont {F.}~\bibnamefont {Corredor}}, \bibinfo {author}
  {\bibfnamefont {R.~D.}\ \bibnamefont {McDonald}}, \bibinfo {author}
  {\bibfnamefont {Z.~X.}\ \bibnamefont {Shen}}, \ and\ \bibinfo {author}
  {\bibfnamefont {I.~R.}\ \bibnamefont {Fisher}},\ }\href {\doibase
  10.1103/PhysRevB.81.205407} {\bibfield  {journal} {\bibinfo  {journal} {Phys.
  Rev. B}\ }\textbf {\bibinfo {volume} {81}},\ \bibinfo {pages} {205407}
  (\bibinfo {year} {2010})}\BibitemShut {NoStop}%
\bibitem [{\citenamefont {Zhang}\ \emph {et~al.}(2010)\citenamefont {Zhang},
  \citenamefont {He}, \citenamefont {Chang}, \citenamefont {Song},
  \citenamefont {Wang}, \citenamefont {Chen}, \citenamefont {Jia},
  \citenamefont {Fang}, \citenamefont {Dai}, \citenamefont {Shan},
  \citenamefont {Shen}, \citenamefont {Niu}, \citenamefont {Qi}, \citenamefont
  {Zhang}, \citenamefont {Ma},\ and\ \citenamefont {Xue}}]{Zhang10}%
  \BibitemOpen
  \bibfield  {author} {\bibinfo {author} {\bibfnamefont {Y.}~\bibnamefont
  {Zhang}}, \bibinfo {author} {\bibfnamefont {K.}~\bibnamefont {He}}, \bibinfo
  {author} {\bibfnamefont {C.-Z.}\ \bibnamefont {Chang}}, \bibinfo {author}
  {\bibfnamefont {C.-L.}\ \bibnamefont {Song}}, \bibinfo {author}
  {\bibfnamefont {L.-L.}\ \bibnamefont {Wang}}, \bibinfo {author}
  {\bibfnamefont {X.}~\bibnamefont {Chen}}, \bibinfo {author} {\bibfnamefont
  {J.-F.}\ \bibnamefont {Jia}}, \bibinfo {author} {\bibfnamefont
  {Z.}~\bibnamefont {Fang}}, \bibinfo {author} {\bibfnamefont {X.}~\bibnamefont
  {Dai}}, \bibinfo {author} {\bibfnamefont {W.-Y.}\ \bibnamefont {Shan}},
  \bibinfo {author} {\bibfnamefont {S.-Q.}\ \bibnamefont {Shen}}, \bibinfo
  {author} {\bibfnamefont {Q.}~\bibnamefont {Niu}}, \bibinfo {author}
  {\bibfnamefont {X.-L.}\ \bibnamefont {Qi}}, \bibinfo {author} {\bibfnamefont
  {S.-C.}\ \bibnamefont {Zhang}}, \bibinfo {author} {\bibfnamefont {X.-C.}\
  \bibnamefont {Ma}}, \ and\ \bibinfo {author} {\bibfnamefont {Q.-K.}\
  \bibnamefont {Xue}},\ }\href {\doibase 10.1038/nphys1689} {\bibfield
  {journal} {\bibinfo  {journal} {Nat. Phys.}\ }\textbf {\bibinfo {volume}
  {6}},\ \bibinfo {pages} {584} (\bibinfo {year} {2010})}\BibitemShut {NoStop}%
\bibitem [{\citenamefont {Pan}\ \emph {et~al.}(2011)\citenamefont {Pan},
  \citenamefont {Vescovo}, \citenamefont {Fedorov}, \citenamefont {Gardner},
  \citenamefont {Lee}, \citenamefont {Chu}, \citenamefont {Gu},\ and\
  \citenamefont {Valla}}]{Pan11}%
  \BibitemOpen
  \bibfield  {author} {\bibinfo {author} {\bibfnamefont {Z.-H.}\ \bibnamefont
  {Pan}}, \bibinfo {author} {\bibfnamefont {E.}~\bibnamefont {Vescovo}},
  \bibinfo {author} {\bibfnamefont {A.~V.}\ \bibnamefont {Fedorov}}, \bibinfo
  {author} {\bibfnamefont {D.}~\bibnamefont {Gardner}}, \bibinfo {author}
  {\bibfnamefont {Y.~S.}\ \bibnamefont {Lee}}, \bibinfo {author} {\bibfnamefont
  {S.}~\bibnamefont {Chu}}, \bibinfo {author} {\bibfnamefont {G.~D.}\
  \bibnamefont {Gu}}, \ and\ \bibinfo {author} {\bibfnamefont {T.}~\bibnamefont
  {Valla}},\ }\href {\doibase 10.1103/PhysRevLett.106.257004} {\bibfield
  {journal} {\bibinfo  {journal} {Phys. Rev. Lett.}\ }\textbf {\bibinfo
  {volume} {106}},\ \bibinfo {pages} {257004} (\bibinfo {year}
  {2011})}\BibitemShut {NoStop}%
\bibitem [{\citenamefont {Zhang}\ \emph {et~al.}(2009)\citenamefont {Zhang},
  \citenamefont {Liu}, \citenamefont {Qi}, \citenamefont {Dai}, \citenamefont
  {Fang},\ and\ \citenamefont {Zhang}}]{Zhang09}%
  \BibitemOpen
  \bibfield  {author} {\bibinfo {author} {\bibfnamefont {H.}~\bibnamefont
  {Zhang}}, \bibinfo {author} {\bibfnamefont {C.-X.}\ \bibnamefont {Liu}},
  \bibinfo {author} {\bibfnamefont {X.-L.}\ \bibnamefont {Qi}}, \bibinfo
  {author} {\bibfnamefont {X.}~\bibnamefont {Dai}}, \bibinfo {author}
  {\bibfnamefont {Z.}~\bibnamefont {Fang}}, \ and\ \bibinfo {author}
  {\bibfnamefont {S.-C.}\ \bibnamefont {Zhang}},\ }\href {\doibase
  10.1038/nphys1270} {\bibfield  {journal} {\bibinfo  {journal} {Nat. Phys.}\
  }\textbf {\bibinfo {volume} {5}},\ \bibinfo {pages} {438} (\bibinfo {year}
  {2009})}\BibitemShut {NoStop}%
\bibitem [{\citenamefont {Liu}\ \emph {et~al.}(2010)\citenamefont {Liu},
  \citenamefont {Qi}, \citenamefont {Zhang}, \citenamefont {Dai}, \citenamefont
  {Fang},\ and\ \citenamefont {Zhang}}]{Liu10}%
  \BibitemOpen
  \bibfield  {author} {\bibinfo {author} {\bibfnamefont {C.-X.}\ \bibnamefont
  {Liu}}, \bibinfo {author} {\bibfnamefont {X.-L.}\ \bibnamefont {Qi}},
  \bibinfo {author} {\bibfnamefont {H.}~\bibnamefont {Zhang}}, \bibinfo
  {author} {\bibfnamefont {X.}~\bibnamefont {Dai}}, \bibinfo {author}
  {\bibfnamefont {Z.}~\bibnamefont {Fang}}, \ and\ \bibinfo {author}
  {\bibfnamefont {S.-C.}\ \bibnamefont {Zhang}},\ }\href {\doibase
  10.1103/PhysRevB.82.045122} {\bibfield  {journal} {\bibinfo  {journal} {Phys.
  Rev. B}\ }\textbf {\bibinfo {volume} {82}},\ \bibinfo {pages} {045122}
  (\bibinfo {year} {2010})}\BibitemShut {NoStop}%
\bibitem [{\citenamefont {Wray}\ \emph {et~al.}(2011)\citenamefont {Wray},
  \citenamefont {Xu}, \citenamefont {Xia}, \citenamefont {Hsieh}, \citenamefont
  {Fedorov}, \citenamefont {Hor}, \citenamefont {Cava}, \citenamefont {Bansil},
  \citenamefont {Lin},\ and\ \citenamefont {Hasan}}]{Wray11}%
  \BibitemOpen
  \bibfield  {author} {\bibinfo {author} {\bibfnamefont {L.~A.}\ \bibnamefont
  {Wray}}, \bibinfo {author} {\bibfnamefont {S.-Y.}\ \bibnamefont {Xu}},
  \bibinfo {author} {\bibfnamefont {Y.}~\bibnamefont {Xia}}, \bibinfo {author}
  {\bibfnamefont {D.}~\bibnamefont {Hsieh}}, \bibinfo {author} {\bibfnamefont
  {A.~V.}\ \bibnamefont {Fedorov}}, \bibinfo {author} {\bibfnamefont {Y.~S.}\
  \bibnamefont {Hor}}, \bibinfo {author} {\bibfnamefont {R.~J.}\ \bibnamefont
  {Cava}}, \bibinfo {author} {\bibfnamefont {A.}~\bibnamefont {Bansil}},
  \bibinfo {author} {\bibfnamefont {H.}~\bibnamefont {Lin}}, \ and\ \bibinfo
  {author} {\bibfnamefont {M.~Z.}\ \bibnamefont {Hasan}},\ }\href {\doibase
  10.1038/nphys1838} {\bibfield  {journal} {\bibinfo  {journal} {Nat. Phys.}\
  }\textbf {\bibinfo {volume} {7}},\ \bibinfo {pages} {32} (\bibinfo {year}
  {2011})}\BibitemShut {NoStop}%
\bibitem [{\citenamefont {Zhou}\ \emph {et~al.}(2017)\citenamefont {Zhou},
  \citenamefont {Jiang}, \citenamefont {Xie},\ and\ \citenamefont
  {Sun}}]{Zhou17}%
  \BibitemOpen
  \bibfield  {author} {\bibinfo {author} {\bibfnamefont {Y.-F.}\ \bibnamefont
  {Zhou}}, \bibinfo {author} {\bibfnamefont {H.}~\bibnamefont {Jiang}},
  \bibinfo {author} {\bibfnamefont {X.~C.}\ \bibnamefont {Xie}}, \ and\
  \bibinfo {author} {\bibfnamefont {Q.-F.}\ \bibnamefont {Sun}},\ }\href
  {\doibase 10.1103/PhysRevB.95.245137} {\bibfield  {journal} {\bibinfo
  {journal} {Phys. Rev. B}\ }\textbf {\bibinfo {volume} {95}},\ \bibinfo
  {pages} {245137} (\bibinfo {year} {2017})}\BibitemShut {NoStop}%
\bibitem [{\citenamefont {Chen}(2020)}]{Chen20_TI_Edelstein}%
  \BibitemOpen
  \bibfield  {author} {\bibinfo {author} {\bibfnamefont {W.}~\bibnamefont
  {Chen}},\ }\href {\doibase 10.1088/1361-648X/ab46c6} {\bibfield  {journal}
  {\bibinfo  {journal} {J. Phys. Condens. Matter}\ }\textbf {\bibinfo {volume}
  {32}},\ \bibinfo {pages} {035809} (\bibinfo {year} {2020})}\BibitemShut
  {NoStop}%
\bibitem [{\citenamefont {Zegarra}\ \emph {et~al.}()\citenamefont {Zegarra},
  \citenamefont {Egues},\ and\ \citenamefont {Chen}}]{Zegarra20}%
  \BibitemOpen
  \bibfield  {author} {\bibinfo {author} {\bibfnamefont {A.}~\bibnamefont
  {Zegarra}}, \bibinfo {author} {\bibfnamefont {J.~C.}\ \bibnamefont {Egues}},
  \ and\ \bibinfo {author} {\bibfnamefont {W.}~\bibnamefont {Chen}},\
  }\href@noop {} {\bibinfo  {journal} {arXiv:2001.01081}\ }\BibitemShut
  {NoStop}%
\end{thebibliography}%

\end{document}